\documentclass[%
 aip,
 amsmath,amssymb,
preprint,%
]{revtex4-2}

\usepackage{graphicx}% Include figure files
\usepackage{dcolumn}% Align table columns on decimal point
\usepackage{bm}% bold math
\usepackage{amsmath}
\usepackage[utf8]{inputenc}
\usepackage[T1]{fontenc}
\usepackage{mathptmx}
\usepackage[colorlinks=true,linkcolor=blue]{hyperref}%
\usepackage{color}
\usepackage{CJK}
\usepackage{float}
\UseRawInputEncoding

\usepackage{etoolbox}

\makeatletter
\def\@email#1#2{%
 \endgroup
 \patchcmd{\titleblock@produce}
  {\frontmatter@RRAPformat}
  {\frontmatter@RRAPformat{\produce@RRAP{*#1\href{mailto:#2}{#2}}}\frontmatter@RRAPformat}
  {}{}
}%
\makeatother

\begin{document}

\preprint{AIP/123-QED}

\begin{CJK*}{UTF8}{gbsn}

\title{Temporally sparse data assimilation for the small-scale reconstruction of turbulence}% Force line breaks with \\
\author{Yunpeng Wang (王云朋)$^{1,2,3}$}
\author{Zelong Yuan (袁泽龙)$^{1,2,3}$}
\author{Chenyue Xie (谢晨月)$^{4}$}
\author{Jianchun Wang (王建春)$^{1,2,3,*}$}
\email[Author to whom correspondence should be addressed:\ ]{wangjc@sustech.edu.cn}
\affiliation{\small 1. National Center for Applied Mathematics Shenzhen (NCAMS), Southern University of Science and Technology, Shenzhen, 518055 China.\\
2. Department of Mechanics and Aerospace Engineering, Southern University of Science and Technology, Shenzhen, 518055 China.\\
3. Guangdong–Hong Kong–Macao Joint Laboratory for Data-Driven Fluid Mechanics and Engineering Applications, Southern University of Science and Technology, Shenzhen, 518055 China.\\
4. Department of Ocean Science, The Hong Kong University of Science and Technology, Hong Kong, 999077 China.}
\date{\today}% It is always \today, today,

\begin{abstract}
Previous works have shown that the small-scale information of incompressible homogeneous isotropic turbulence (HIT) is fully recoverable as long as sufficient large-scale structures are continuously enforced through temporally continuous data assimilation (TCDA). In the current work, we show that the assimilation time step can be relaxed to values about 1 $\sim$ 2 orders larger than that for TCDA, using a temporally sparse data assimilation (TSDA) strategy, while the accuracy is still maintained or even slightly better in the presence of non-negligible large-scale errors. The one-step data assimilation (ODA) is examined to unravel the mechanism of TSDA. It is shown that the relaxation effect for errors above the assimilation wavenumber $k_a$ is responsible for the error decay in ODA. Meanwhile, The errors contained in the large scales can propagate into small scales and make the high-wavenumber ($k>k_a$) error noise decay slower with TCDA than TSDA. This mechanism is further confirmed by incorporating different levels of errors in the large scales of the reference flow field. The advantage of TSDA is found to grow with the magnitude of the incorporated errors. Thus, it is potentially more beneficial to adopt TSDA if the reference data contains non-negligible errors. Finally, an outstanding issue raised in previous works regarding the possibility of recovering the dynamics of sub-Kolmogorov scales using direct numerical simulation (DNS) data at Kolmogorov scale resolution is also discussed.

%\begin{description}
%\item[PACS number(s)]
%47.20.Ky, 47.20.Lz, 47.20.Ma
%\end{description}
\end{abstract}

\pacs{Valid PACS appear here}% PACS, the Physics and Astronomy
\maketitle
\end{CJK*}

%\tableofcontents

\section{\label{sec:level1}Introduction}

Accurate prediction of turbulent flows is crucial in many areas of research and engineering community, including meteorology, aerospace engineering, oceanology, geosciences, environmental and industrial applications. \cite{Charney1969,Olson2003,Meldi2017} Among various kinds of flow prediction methods, computational fluid dynamics (CFD) has become a major tool in recent decades due to the significant developments in modern computers and numerical methods. \cite{Kato2011} However, due to the strong sensitivity of turbulence to small perturbations, \cite{Leith1972} different CFD realizations with small differences in the initial conditions can go through diverged trajectories in phase space. Indeed, real world measurement inevitably contains errors, making it impossible to exactly prescribe the initial conditions for a CFD solver. Meanwhile, CFD simulations always contain numerical errors, which probably further contaminate the CFD-based solution. On the other hand, however, one may have a time sequence of observational data, which may be sparse in time and/or space. These observational data also contains some information on the true state of the flow field. \cite{Bauweraerts2021} In this case, improved predictions of the flow field can be expected if the CFD simulation and the observational data can be properly integrated. Such a strategy belongs to a well established independent subject, namely the data assimilation (DA) of dynamical system. \cite{Lewis2006}

The commonly adopted methodology of DA for fluid mechanics include temporally continuous data assimilation (TCDA) through direct data embedding, \cite{Olson2003,Charney1969,Yoshida2005,Lalescu2013} spatial and Fourier nudging, \cite{Buzzicotti2020,ClarkDiLeoni2020,Pawar2020} Kalman filtering-based methods, \cite{Meldi2017,Colburn2011,Suzuki2012,Kato2015,Deng2021,Strofer2021,Fang2021,Wang2021} adjoint-based variational methods \cite{Bewley2004,Gronskis2013,Foures2014,Mons2017,He2019,Chen2019,He2020a,He2020b,Li2022,He2022} and forward sensitivity method. \cite{Lakshmivarahan2010} Recently, due to the prosperity of the machine-learning techniques, \cite{Ling2016,Wang2017,Maulik2017,Duraisamy2019,Yu2019,Srinivasan2019,Ma2019,Xie2020,Yuan2020,Dominique2022} artificial neural networks have also become powerful tools for DA. \cite{Sirignano2020,Deng2021,Xu2021,Arzani2021} 

In the direct numerical simulation (DNS), DA is widely exploited in problems including the estimation of initial and boundary conditions, \cite{Gronskis2013} the exact flow field reconstruction using large-scale information \cite{Olson2003,Yoshida2005,Lalescu2013} and data compression. \cite{Wu2020} In the Reynolds averaged Navier-Stokes (RANS) simulations and large-eddy simulations (LES), \cite{Sagaut2006} DA is mainly used to calibrate the unclosed Reynolds stress or the subgrid-scale (SGS) stress by either directly representing the whole, or part of, the unclosed terms \emph{per se}, \cite{Foures2014,He2021,Sirignano2020} or by tuning the model parameters. \cite{He2018,Kato2015,Li2017} Other applications of DA in fluid mechanics can be found in the areas such as the optimization of sensor locations, \cite{Mons2017,Akhtar2015,Deng2021} flow controls, \cite{Gunzburger2000,Greenblatt2006,Marquet2008} flow field estimation \cite{Bewley2004,He2020a,Colburn2011,Suzuki2012,Zhang2021,Wang2022,Agasthya2022} and parameter estimation for wind tunnel wall interference corrections. \cite{Ma2016,Belligoli2021}

Of closer relevance to the current study is the exact small-scale flow reconstruction of incompressible homogeneous isotropic turbulence (HIT) through direct data embedding in the Fourier space. \cite{Olson2003,Yoshida2005,Lalescu2013} Even though turbulence has a strong sensitivity to small perturbations in the initial condition, \cite{Leith1972} existing works have shown that this sensitivity can be overcome if sufficient amount of large-scale structures in a flow field are continuously enforced with time. \cite{Yoshida2005,Lalescu2013} More precisely, the small-scale information of a flow is completely slaved to the Fourier modes above a critical length scale of wavenumber $k_c$. Slightly different thresholds for $k_c$ have been reported, such as $k_c \eta \approx 0.2$ \cite{Yoshida2005} and $k_c \eta \approx 0.15$, \cite{Lalescu2013} where $\eta$ is the Kolmogorov length scale. Hence, the spatial-temporal information of a flow can be completely recovered with machine-error level accuracy, as long as the dominant large-scale modes are continuously known and enforced at the same temporal resolution of the reference DNS.

In the present work, we shall further show that,  even though the number of necessary large spatial scales for exact flow reconstruction by DA is fixed, the time step for DA can be relaxed to much larger values compared to the temporally continuous data assimilation (TCDA). Meanwhile, the accuracy is still maintained at a similar level or even slightly better in the presence of non-negligible large-scale errors. In this case, the amount of the required data for the assimilation process can be largely reduced. From another point of view, the current treatment can be viewed as a further data compression in time, in addition to the compression in spatial scales. Nevertheless, the current treatment is essentially different from the recently proposed data compression scheme, \cite{Wu2020} which requires the data of the entire spatial domain at large time steps instead of only the large-scale spatial modes.

As noted in the work by Lalescu \emph{et al.}, \cite{Lalescu2013} another motivation of TCDA is that existing works have shown that length scales smaller than the Kolmogorov length scale $\eta$ can exist due to the spatial intermittency in turbulence. \cite{Anselmet1984,Paladin1987,Yakhot2005,Schumacher2007,Gibbon2012} Consequently, the DNS at a Kolmogorov scale resolution can be quite inaccurate. \cite{Yakhot2005} A counter argument is that the sub-Kolmogorov scales should be completely recoverable from the DNS of Kolmogorov resolution if the small-scale structures are dominated by the large-scale dynamics. However, Lalescu \emph{et al.} \cite{Lalescu2013} have also realized that the large-scale data used in the TCDA process comes from a projection of the solution using a fine grid which is the same as the adopted grid in the DA-based simulation. \cite{Yoshida2005,Lalescu2013} Hence, whether the sub-Kolmogorov-scale recovery is truly viable is still in question, since the grid used for the Kolmogorov scale-based DNS is a `coarser' grid compared to the `ideal' grid with which the sub-Kolmogorov scales can be properly resolved. In this regard, the sub-Kolmogorov-scale recovery is also examined and discussed in the current study.

The rest of the paper is organized as follows. The governing equations of incompressible turbulence and a brief introduction to the DA in the Fourier space are given in Section II, followed by a detailed analysis of TSDA in Section III, where TCDA and TSDA are compared and examined in detail through their total error and error spectra. The results for a higher Reynolds number case is briefly discussed in Section IV, where some slightly different behaviors are observed. In Section V, the possibility of sub-Kolmogorov-scale recovery is examined using both TSDA and TCDA. Finally, a brief summary of the paper and comments on future works are given in Section VI.

\section{Governing equations of incompressible turbulence and a brief introduction of Fourier-space data assimilation}

The governing equations of incompressible turbulence are first presented in the current section. Subsequently, the data assimilation in Fourier space is briefly introduced with some previous findings revisited.

\subsection{Governing equations of incompressible turbulence}

For incompressible turbulence, the mass and momentum conservation are governed by the Navier-Stokes equations, namely: \cite{Pope2000,Ishihara2009}
 \begin{equation}
  \frac{\partial u_{i}}{\partial x_{i}}=0,
  \label{mass}
\end{equation}
 \begin{equation}
  \frac{\partial u_{i}}{\partial t}+\frac{\partial u_{i} u_{j}}{\partial x_{j}}=-\frac{\partial p}{\partial x_{i}}+\nu\frac{\partial^{2} u_{i}}{\partial x_{j}\partial x_{j}}+F_{i},
  \label{momentum}
\end{equation}
where $u_{i}$ is the velocity component in the $i$ coordinate direction, $\nu$ is the kinematic viscosity, $p$ is the pressure divided by the constant density (i.e. in units of $u^2$), and $F_i$ is the large-scale forcing applied to the two lowest wavenumber shells. \cite{Wang2010,Wang2018} The summation convention is used throughout the paper unless otherwise noted. In the current work, these equations are used in dimensionless forms with $\nu=1/Re$ being the reciprocal of the reference Reynolds number $Re$, defined as $Re=u_f L_f/\nu_f$. Here $u_f$ is the reference velocity, $L_f$ is the reference length, and $\nu_f$ is the reference kinematic viscosity. In this case, the velocity, spatial coordinate and time are normalized by $u_f$, $L_f$ and $L_f/u_f$ respectively. In addition, pressure is normalized by $\rho_f u_f^2$ with $\rho_f$ being the reference density.

The Kolmogorov length scale $\eta$ is defined by \cite{Pope2000}
\begin{equation}
  \eta=(\frac{\nu^{3}}{\epsilon})^{1/4},
  \label{eta}
\end{equation}
where $\epsilon$ is the spatially averaged dissipation rate given by $\epsilon=2\nu\langle S_{ij}S_{ij}\rangle$ with $S_{ij}=\frac{1}{2}(\partial{u_{i}}/\partial{x_{j}}+\partial{u_{j}}/\partial{x_{i}})$
being the strain rate tensor. In the current work, $\langle \cdot \rangle$ invariably stands for the spatial average over the entire physical domain. The Kolmogorov length scale quantifies the size of the smallest eddies in turbulence. Correspondingly, the Kolmogorov time scale is calculated as
\begin{equation}
  \tau_{\eta}=(\frac{\nu}{\epsilon})^{1/2}.
  \label{taueta}
\end{equation}
In addition, the Taylor length scale $\lambda$ is defined by
\begin{equation}
  \lambda=\sqrt{\frac{5\nu}{\epsilon}}u^{rms},
  \label{lambda}
\end{equation}
with $u^{rms}=\sqrt{\langle u_{i}u_{i}\rangle}$ being the root-mean-square (rms) value of velocity magnitude. The Reynolds number based on the Taylor length scale can be calculated as \cite{Ishihara2009}
\begin{equation}
  Re_{\lambda}=\frac{u^{rms}\lambda}{\sqrt{3}\nu}.
  \label{ReMt}
\end{equation}
Finally, the kinetic energy per unit mass is given by
\begin{equation}
  \int_{0}^{\infty}E(k)dk=\frac{(u^{rms})^{2}} {2},
  \label{Ek}
\end{equation}
where $E(k)$ is the energy spectrum.

\subsection{The data assimilation in Fourier space}

The sensitivity of turbulence to small errors in the initial conditions is well acknowledged in the community of fluid dynamics. \cite{Leith1972} Consequently, the predictability of turbulence in real world applications is quite limited. In spite of this sensitivity, several works \cite{Olson2003,Yoshida2005,Lalescu2013} have shown that, under certain conditions, the perturbation errors can be gradually erased with time if the correct large-scale Fourier modes are continuously supplied to the numerical solution. In other words, the growth of errors due to the aforementioned sensitivity can be gradually suppressed by the dominating effect of large-scales dynamics over the small-scale dynamics in the cost of a continuous time sequence of large-scale data. The corresponding data assimilation procedure in Fourier space is briefly introduced in the following.

First, we assume that there are two time sequences of numerical solutions of Eqs.~(\ref{mass}) and (\ref{momentum}), denoted by $\mathbf{u}$ and $\mathbf{u}^{ref}$, computed using the same temporal and spatial resolutions but different initial conditions. Here, we let $\mathbf{u}^{ref}$ denote the true solution and $\mathbf{u}$ be the solution with errors induced by the deviation from the true initial condition (i.e. the initial condition used by $\mathbf{u}^{ref}$). Previous works have found that the solution of $\mathbf{u}$ would converge to $\mathbf{u}^{ref}$ as long as the Fourier modes of $\mathbf{u}$ above a critical length scale are continuously replaced by the corresponding values of $\mathbf{u}^{ref}$ as the numerical solution marches forward. \cite{Yoshida2005} The threshold can be written as $k_a>k_c$, where $k_a$ is the assimilation wavenumber below which all the Fourier modes of $\mathbf{u}$ are replaced by that of $\mathbf{u}^{ref}$. $k_c$ is the critical wavenumber necessary for the success of such an assimilation, and it depends on the Kolmogorov length scale $\eta$. By letting $\mathbf{\hat{u}}(\mathbf{k},t)$ denote the Fourier coefficient of $\mathbf{u}$ at wavenumber $\mathbf{k}$ and time $t$, the assimilation procedure can be written as

\begin{equation}
  \mathbf{\hat u}(\mathbf{k},t_0+n \Delta T)=\mathbf{\hat u}^{ref}(\mathbf{k},t_0+n \Delta T), \text{for}~k<k_a,
  \label{assi}
\end{equation}
where $\Delta T$ is the discrete time step for the data assimilation procedure, $n=1,2,3\ldots$ and $k$ is the magnitude of the wavenumber vector. In the continuous limit, $\Delta T=\Delta t$, with $\Delta t$ being the time step of DNS. Different thresholds for $k_c$ in the continuous limit have been reported, namely $k_c \eta=0.2$ \cite{Yoshida2005} and $k_c \eta=0.15$. \cite{Lalescu2013} In the current work, we shall show that even though the spatial threshold in terms of the necessary Fourier modes is fixed, the size of the assimilation time step $\Delta T$ required for the convergence of $\mathbf{u}$ to $\mathbf{u}^{ref}$ can be relaxed to a much larger value compared to a continuous assimilation.

\section{Temporally continuous and sparse data assimilation of incompressible isotropic turbulence}

In the present study, the data assimilation is implemented through direct numerical simulation (DNS) of a forced incompressible isotropic turbulence. The DNSs are performed in a cubic box of $(2\pi)^{3}$ with periodic boundary conditions and a uniform grid spacing denoted by $h_{DNS}$. In this case, the pseudospectral approach is conveniently adopted.\cite{Ishihara2009} Meanwhile, the second-order two-step Adams-Bashforth scheme is used for time marching. The velocity field is forced by prescribing the energy spectrum within the two lowest wavenumber shells. More precisely, $E(1)=1.242477$ and $E(2)=0.391356$ which closely follow the $-5/3$ rule. \cite{Wang2010} Full dealiasing is implemented using the two-thirds rule, \cite{Patterson1971} with the maximum resolved wavenumber given by $k_{max}=N/3$, where $N$ is the number of grid points in each spatial direction. For each grid resolution, a preceding simulation is first run until a fully-developed statistically stationary state is reached. Subsequently, the initial conditions for all the data assimilation experiments are obtained from the fully-developed flow field.

In the current section, three simulation cases for $\mathbf{u}^{ref}$ are tested, whose simulation parameters are listed in Table~\ref{tab1}. As shown in previous work, \cite{Yoshida2005} the influences of different choices of resolution parameters $k_{max} \eta$ on the results of the data assimilation are quite small. Consequently, following the previous work, \cite{Yoshida2005} the resolution parameters are all chosen such that $k_{max} \eta \approx 1$ in the present section. Nevertheless, a test on the influence of a larger resolution parameter (i.e. finer grid) is given in Appendix A for interested readers. It is worth emphasizing that the time step $\Delta t$ is much smaller than the Kolmogorov time scale $\tau_{\eta}$ due to the CFL condition for numerical stability. \cite{Wang2010} In general, $\tau_{\eta} \approx 10$ to $100 \ \Delta t$. In the current analysis, the `erroneous' field $\mathbf{u}$ is generated by adding a perturbation to the initial condition, namely
\begin{equation}
  \mathbf{\hat u}(\mathbf{k},t_0)=(1+\varepsilon)\mathbf{\hat u}^{ref}(\mathbf{k},t_0),
  \label{perturb}
\end{equation}
where $\varepsilon$ is a small real number. This represents a perturbation in the magnitude of the Fourier modes. In all the tested cases of the current section, $\varepsilon=10^{-2}$ is consistently adopted such that both the decay of error in a successful assimilation and the growth of error in an unsuccessful assimilation can be clearly observed. On the contrary, one may possibly not observe the error growth if initial perturbation was too large, or the error decay if the perturbation was too small (e.g. close to the machine-error level). Otherwise, the way how the initial error is imposed does not affect any of the results based on our test. For completeness, some results for the cases in which the initial field is perturbed by a small phase shift in the Fourier modes are briefly presented in Appendix B.

\begin{table*}
\begin{center}
\small
\begin{tabular*}{0.8\textwidth}{@{\extracolsep{\fill}}cccccccc}
\hline
Reso. &$Re_{\lambda}$ &$\eta$ &$k_{max} \eta$ &$ \tau_{\eta}$ &$\nu$ &$\Delta t$ &$\tau_{\eta} / \Delta t$\\ \hline
$64^3$ &60 &0.05 &1.07 &0.165 &0.015 &0.0032 &52\\ 
$128^3$ &105 &0.024 &1.02 &0.093 &0.006 &0.0016 &58\\ 
$256^3$ &160 &0.012 &1.00 &0.057 &0.0024 &0.0008 &71\\ \hline
\end{tabular*}
\normalsize
\caption{Numerical simulation parameters of incompressible isotropic turbulence.}\label{tab1}
\end{center}
\end{table*}

To quantify the assimilation error, we define the magnitude of the error vector in the Fourier space as
\begin{equation}
  \delta=\sqrt{\sum_\mathbf{k} (\mathbf{\hat u}-\mathbf{\hat u}^{ref})^2},
  \label{error}
\end{equation}
where the dependence on time is implicit. Here, we emphasize that Eq.~(\ref{error}) gives equal weight to errors at large and small scales. Considering the disproportionate energy density of large scales, this definition gives relatively more weights to the accuracy of large scales, compared to small scales. Nevertheless, if $\delta$ diminish to the machine-error level (i.e. errors at all scales vanish) as the solution evolves with time, the data assimilation can be deemed as successful. Single-precision solver is adopted in the main body of the work, similar to the work by Lalescu \emph{et al.} \cite{Lalescu2013} In the case of double-precision solver, a brief demonstration is given in Appendix C.

\subsection{The temporally continuous and sparse data assimilation}

\begin{figure}\centering
\includegraphics[width=.34\textwidth]{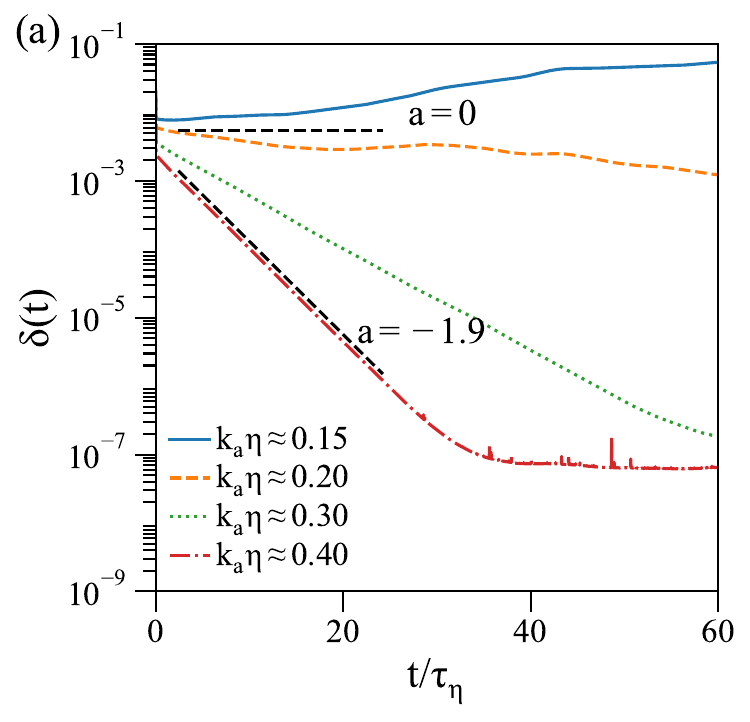}\hspace{-0.10in}
\includegraphics[width=.34\textwidth]{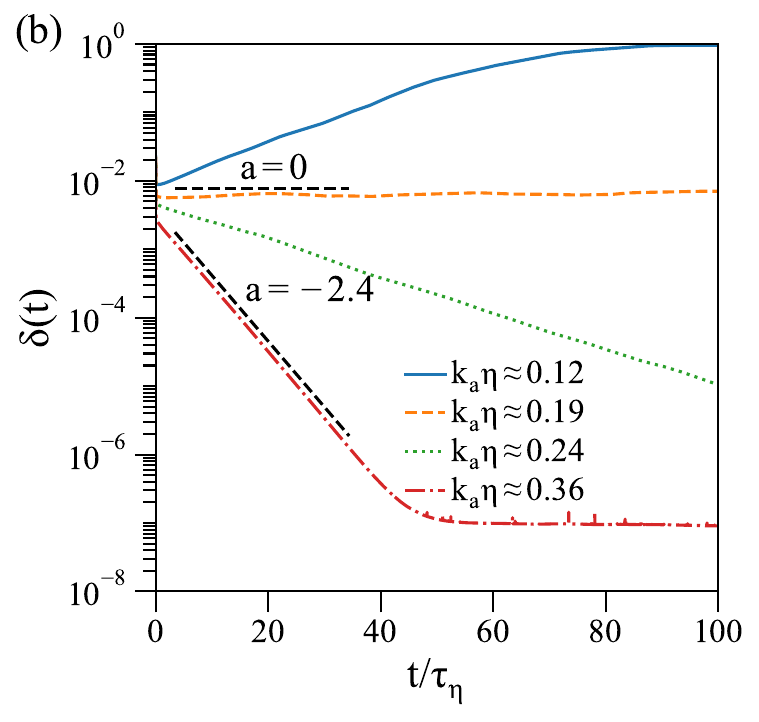}\hspace{-0.10in}
\includegraphics[width=.34\textwidth]{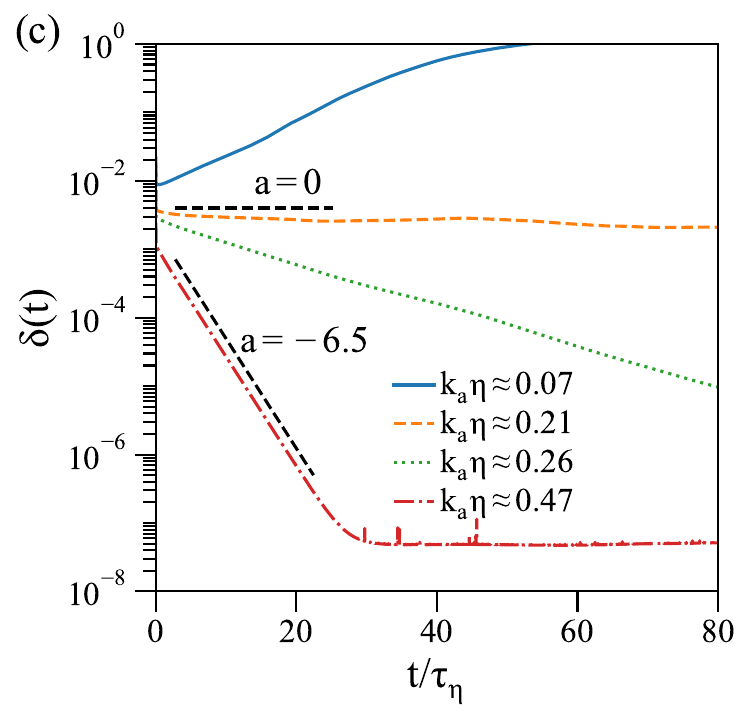}\hspace{-0.10in}
 \caption{The temporal evolution of the error magnitude for the temporally continuous data assimilation: (a) $N=64$, $Re_{\lambda}=60$; (b) $N=128$, $Re_{\lambda}=105$; (c) $N=256$, $Re_{\lambda}=160$.}\label{fig1}
\end{figure}

To briefly revisit the temporally continuous data assimilation (TCDA) reported in previous works, \cite{Yoshida2005,Lalescu2013} we plot the temporal evolution of the error magnitude in Fig.~\ref{fig1}. As previously discussed, the continuous limit is approximated by taking the assimilation time step $\Delta T$ equals to the DNS time step $\Delta t$. Similar to the previous findings, \cite{Yoshida2005,Lalescu2013} the error caused by the difference of the initial condition can be gradually reduced to machine level (i.e. single precision), provided that the assimilation wavenumber $k_a$ is above a critical value $k_c$. Meanwhile, an exponential growth or decay is observed in agreement with previous findings. Here the decay (or growth) constant $a$ is defined through
\begin{equation}
  \delta(t)=\delta_0 e^{a t},
  \label{exponentialgrowth}
\end{equation}
where $\delta_0$ is the initial error magnitude. In Fig.~\ref{fig1}, the values of the exponential constants are also indicated for the steepest decaying curve and the neutral curve ($a=0$).

\begin{figure}\centering
\includegraphics[width=.45\textwidth]{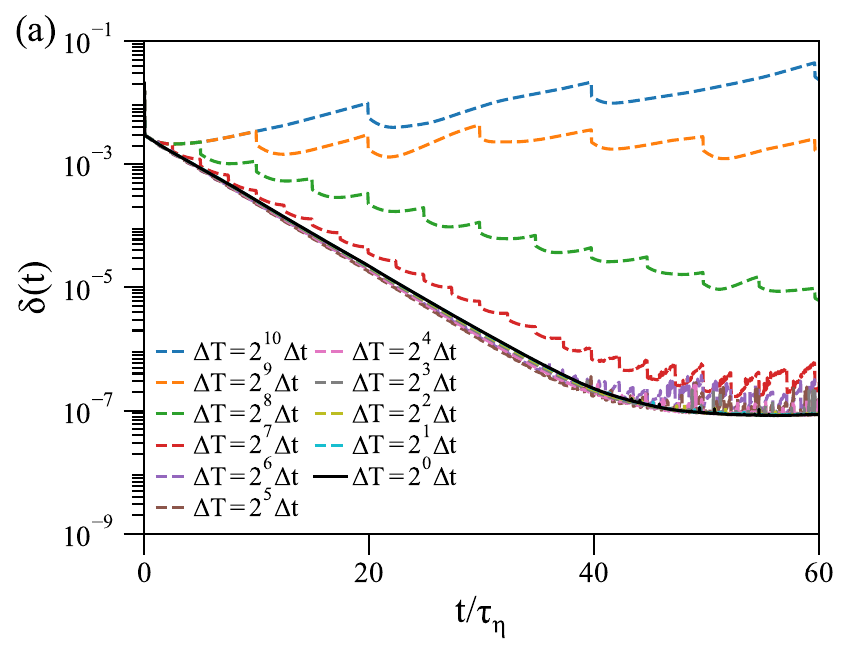}\hspace{-0.10in}
\includegraphics[width=.45\textwidth]{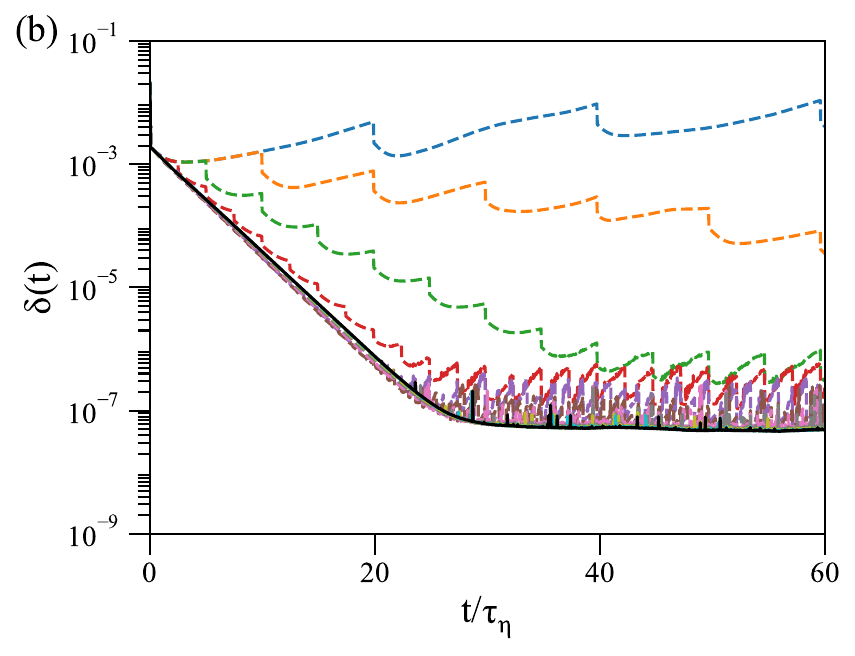}\vspace{-0.10in}

\includegraphics[width=.45\textwidth]{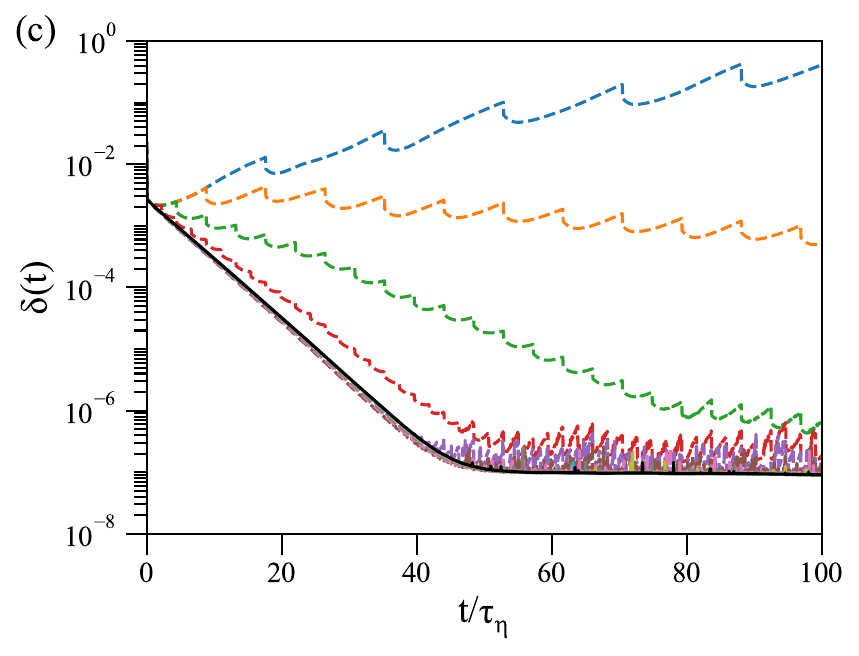}\hspace{-0.10in}
\includegraphics[width=.45\textwidth]{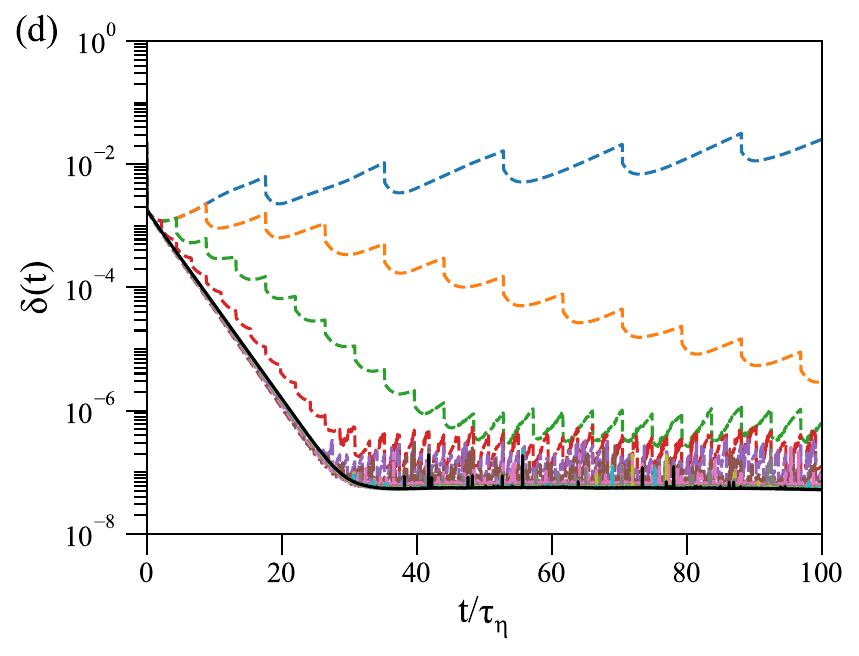}\vspace{-0.10in}

\includegraphics[width=.45\textwidth]{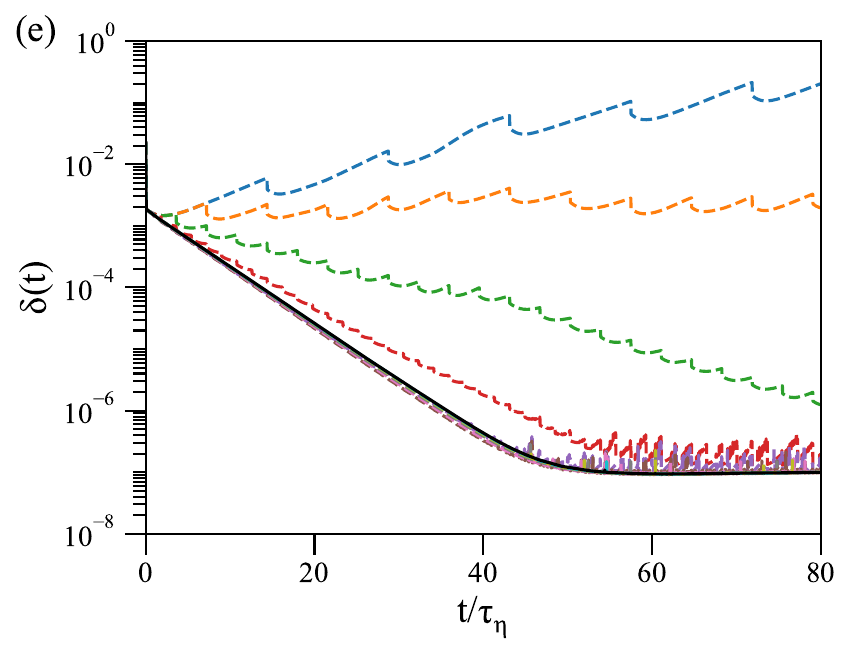}\hspace{-0.10in}
\includegraphics[width=.45\textwidth]{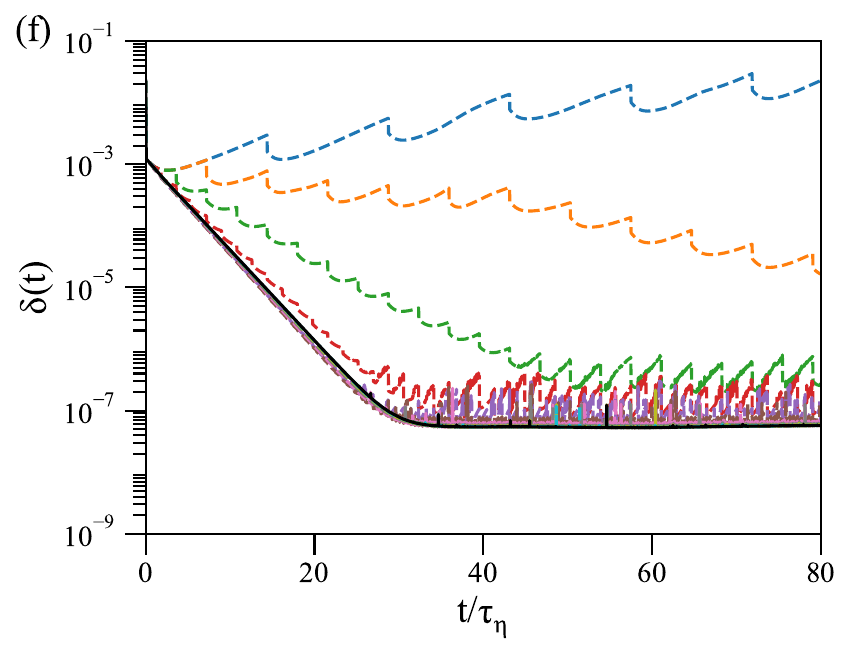}\vspace{-0.10in}
 \caption{The temporal evolution of the error magnitude for the temporally sparse data assimilation: (a) $Re_{\lambda}=60, k_a \eta=0.35$; (b) $Re_{\lambda}=60, k_a \eta=0.45$; (c) $Re_{\lambda}=105, k_a \eta=0.35$; (d) $Re_{\lambda}=105, k_a \eta=0.45$; (e) $Re_{\lambda}=160, k_a \eta=0.35$; (f) $Re_{\lambda}=160, k_a \eta=0.45$.}\label{fig2}
\end{figure}

We also observe in Fig.~\ref{fig1} that, in the $N=64$ case, the exponential behavior is not very concrete for $k_a \eta=0.15$ and $0.20$, presumably because these curves are too close to the neutral state. In the rest of this section, we shall deviate from the continuous limit and explore the possibility of relaxing assimilation time step $\Delta T$ from the DNS value to much larger ones using temporally sparse data assimilation (TSDA), namely

\begin{equation}
  \Delta T= 2^m \Delta t,
  \label{DT}
\end{equation}
where $m=1,2,3\ldots 10$. The evolutions of the magnitude of assimilation error for TSDA are displayed in Fig.~\ref{fig2}. To avoid the inconsistency caused by the initial Euler step used in the two-step Adams-Bashforth scheme, four consecutive (continuous) assimilation steps are performed initially. This should not affect any conclusion since the assimilation can be simply assumed to start from a different initial error field.

In Fig.~\ref{fig2}, the values of $k_a \eta$ for all cases are invariably taken at 0.35 and 0.45 such that the errors in all the TCDAs can reach the machine-error level within reasonable duration. As can be observed in Fig.~\ref{fig2}, these assimilation cases can still be successful with large assimilation steps even though some oscillatory behaviors are present. Surely, one can expect the data assimilation to be still viable with time steps slightly larger than that in the continuous case. However, something more interesting, and to some extent unexpected, is that in many assimilation cases (e.g. $\Delta T/\Delta t \le 2^6$) the assimilation error exhibits the same decaying rate as the continuous one, or even decays slightly faster. As such, the amount of data required in the assimilation process can be significantly reduced while the assimilation accuracy is maintained at the same level of the continuous case.

\begin{figure}\centering
\includegraphics[width=.5\textwidth]{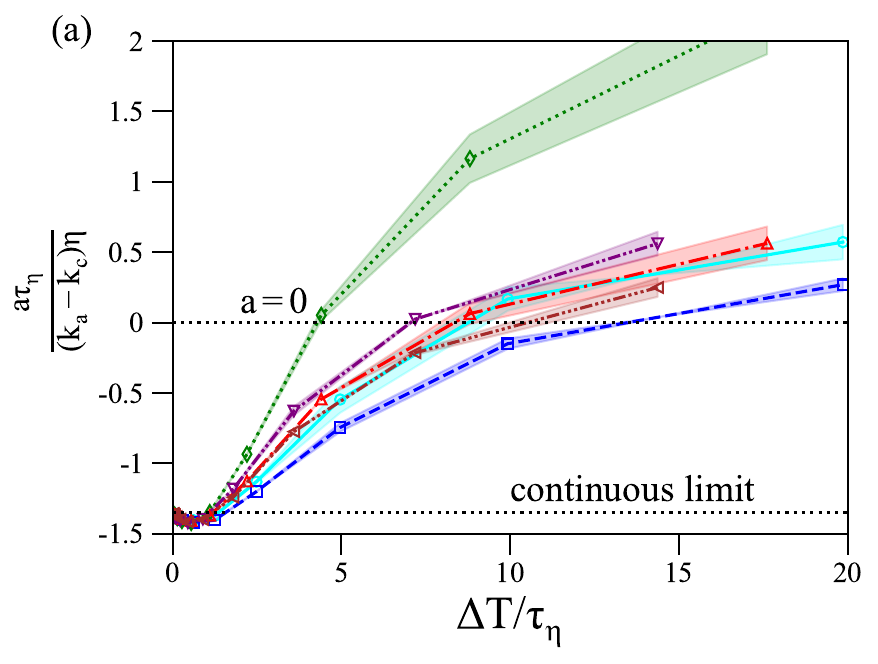}\hspace{-0.10in}
\includegraphics[width=.5\textwidth]{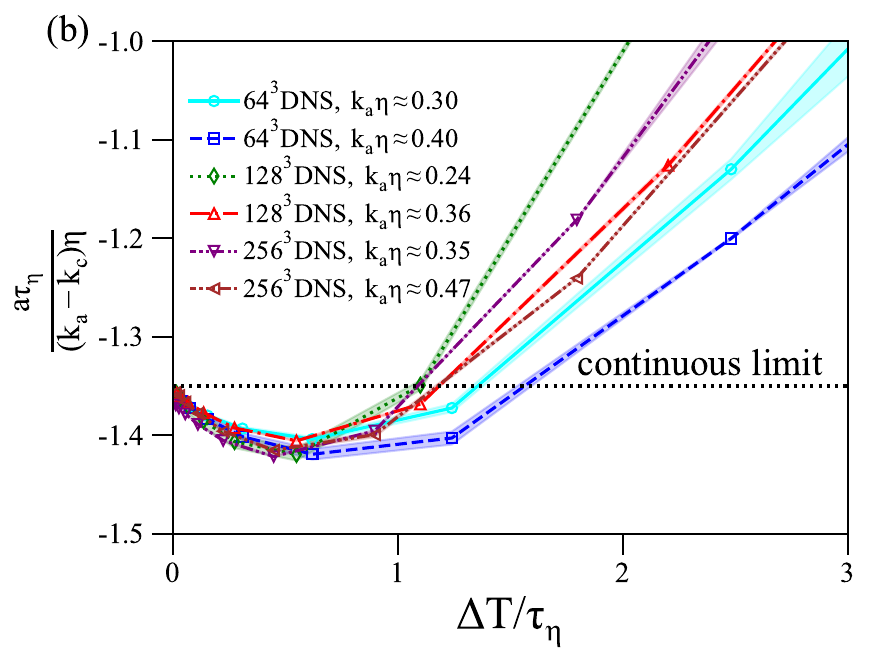}\vspace{-0.10in}
 \caption{The change of the normalized decay constant with respect to the normalized assimilation time interval. Here the shaded area represents the uncertainty when the range of least-square linear fitting is varied: (a) $0<\Delta T<20 \tau_{\eta}$; (b) $0<\Delta T<3 \tau_{\eta}$.}\label{fig3}
\end{figure}

Before further analysis on the assimilation error, we show in Fig.~\ref{fig3} the decay constant of the assimilation error as a function of the assimilation time step $\Delta T$. These decay constants are obtained using different values of $k_a \eta$ through the least-square linear regression method. Here, only the data before the saturation of errors are considered (i.e. only the exponential growing or decaying regions are considered). Even so, the exponential region is still difficult to exactly prescribe for TSDA especially when $\Delta T$ is large. In this case, we additionally use the shaded area in Fig.~\ref{fig3} to represent the uncertainty when the range of the linear fitting is varied: for the $\Delta T \le 128\Delta t$ cases, the exponential behavior is quite concrete, and most of the exponential decaying regions before the error saturation are considered, with the first 1400 to 1500 $\Delta t$ considered for the $N=64$ and $N=256$ cases, and the first 1900 to 2000 $\Delta t$ for the $N=128$ case, based on the respective lengths of the exponential regions for the given values of $k_a \eta$; for the $\Delta T > 128\Delta t$ cases, since there is no well-defined exponential regions, the first 1024 to 2048 $\Delta t$ are considered so that at least two integer multiples of $1024\Delta t$ are considered. It is not surprising that there are more uncertainties with larger assimilation time step $\Delta T$ due to the difficulty in defining the exponential constants. In Fig.~\ref{fig3}, the assimilation time step is normalized by the Kolmogorov time scale, calculated as $\tau_{\eta}=\sqrt{\nu / \epsilon}$, \cite{Pope2000} and the exponential constant is scaled by $(k_a-k_c) \eta / \tau_{\eta}$.

In Fig.~\ref{fig3}a, we observe that, for an assimilation wavenumber that is adequate in the continuous case (i.e. $k_a \eta > k_c \eta$), the assimilation may not be successful if a far too large $\Delta T$ is adopted. In principle, there should also be a critical time step for each of the assimilation wavenumbers, represented by the intersections of the curves and the $a=0$ line. Unfortunately, attempts to re-scale these intersections into a single point turn out unsuccessful based on our tests. However, the curves seem to collapse much better below the continuous limit as shown in Fig.~\ref{fig3}b, which is a zoom-in view of Fig.~\ref{fig3}a in the region close to the continuous limit. This is by no doubt more attractive since it gives a threshold of time scale for TSDA such that its performance is at least the same as, or even slightly better than, TCDA. Consequently, the amount of the required large-scale data of the `true' field $\mathbf{u}^{ref}$ can be tangibly reduced while the accuracy is maintained at a similar level. Figure~\ref{fig3}b shows that as long as the assimilation time step is around $1 \sim 1.5$ times the Kolmogorov time scale, the decaying rate of the error field for TSDA can be as good as TCDA.

As well known, in a typical DNS solution, while the grid space is close to the Kolmogorov length scale $\eta$, the adopted time step is often determined through the CFL condition. \cite{Wang2010} In consequence, the time step of DNS is generally one or two orders smaller than the Kolmogorov time scale (cf. Table~\ref{tab1}). In this sense, the current finding gives a relatively consistent threshold regardless of the adopted time step in DNS. Further, it can be seen from Fig.~\ref{fig3}b that all the curves merge approximately at the same decay rate in the continuous limit. This is in agreement with the finding by Lalescu \emph{et al.}, \cite{Lalescu2013} who found that the decay rates in TCDA follow closely a linear law, namely

\begin{equation}
  a \tau_{\eta} \approx -\beta (k_a - k_c) \eta,
  \label{linearlaw}
\end{equation}
where the constant $-\beta \approx -1.35$ represents the intersection of the curves with the vertical axis.

\begin{figure}\centering
\includegraphics[width=.9\textwidth]{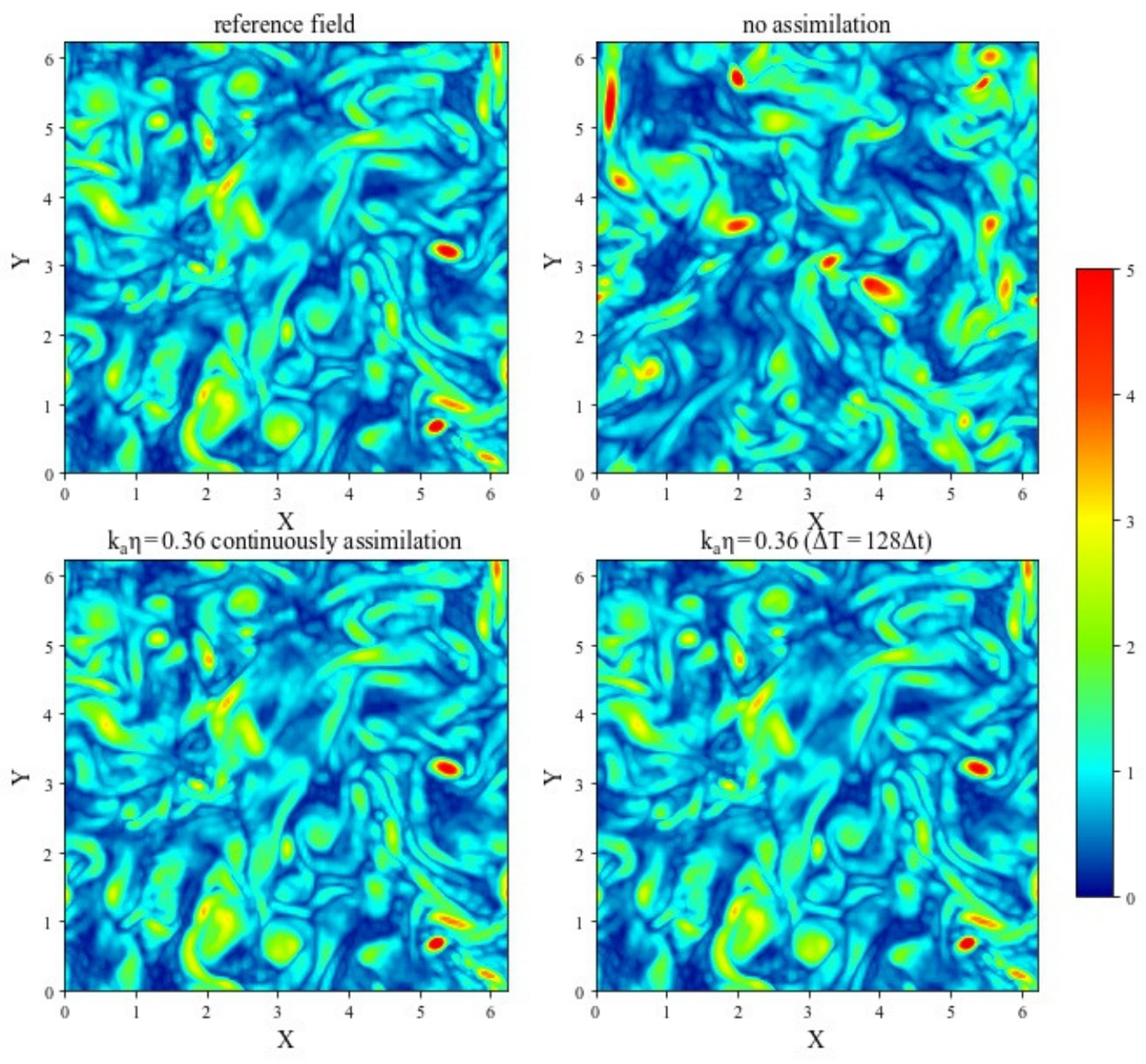}\hspace{-0.10in}
 \caption{The contour field of the normalized vorticity in the temporally continuous and sparse data assimilations at $t \approx 138 \tau_{\eta}$.}\label{fig4}
 \end{figure}

To more directly visualize the performance of the temporally sparse assimilation, we show in Fig.~\ref{fig4} the instantaneous vorticity field at arbitrarily selected $x$-$y$ plane at the end of assimilation ($t \approx 138 \tau_{\eta}$) for the $Re_{\lambda}=105$ case. Here, the vorticity is normalized by its rms value. As can be seen, hardly any similarity in the vorticity field can be recognized between reference field and the one without assimilation due to the uncontrolled growth of the initial error. On the other hand, TCDA gives exactly the same vorticity field as the reference field with machine-level errors unrecognizable from the figure. More importantly, TSDA also recovers the vorticity field correctly even though the time step $\Delta T$ is two orders larger than that for TCDA, demonstrating its great advantage in data requirement and computational efficiency. In the following, a detailed analysis is performed so as to shed some light on the mechanism behind TSDA.

\subsection{The mechanism behind the temporally sparse data assimilation}

\begin{figure}\centering
\includegraphics[width=.34\textwidth]{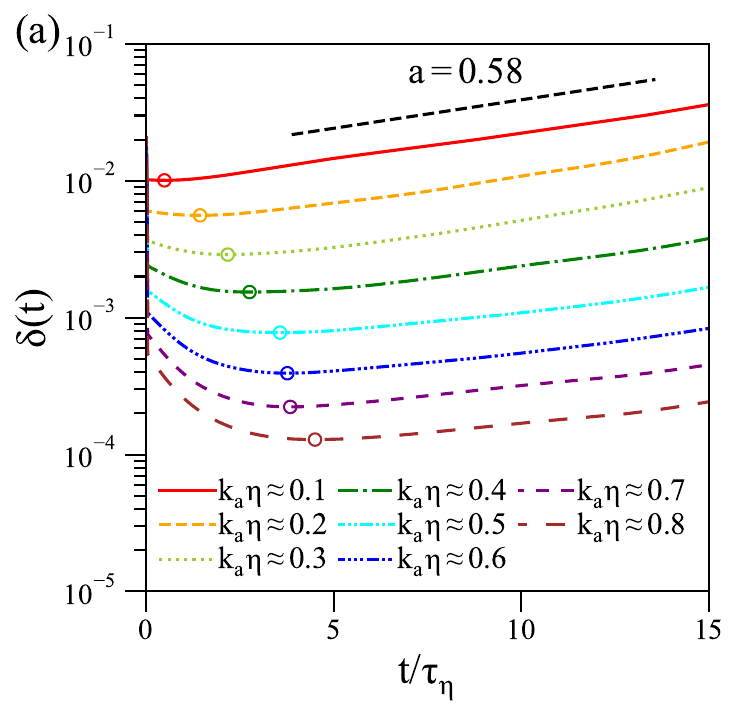}\hspace{-0.10in}
\includegraphics[width=.34\textwidth]{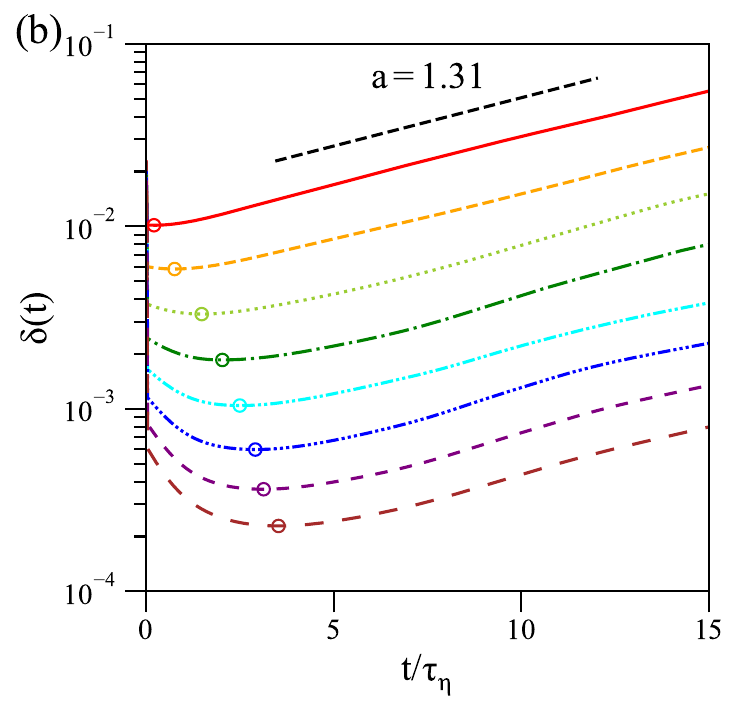}\hspace{-0.10in}
\includegraphics[width=.34\textwidth]{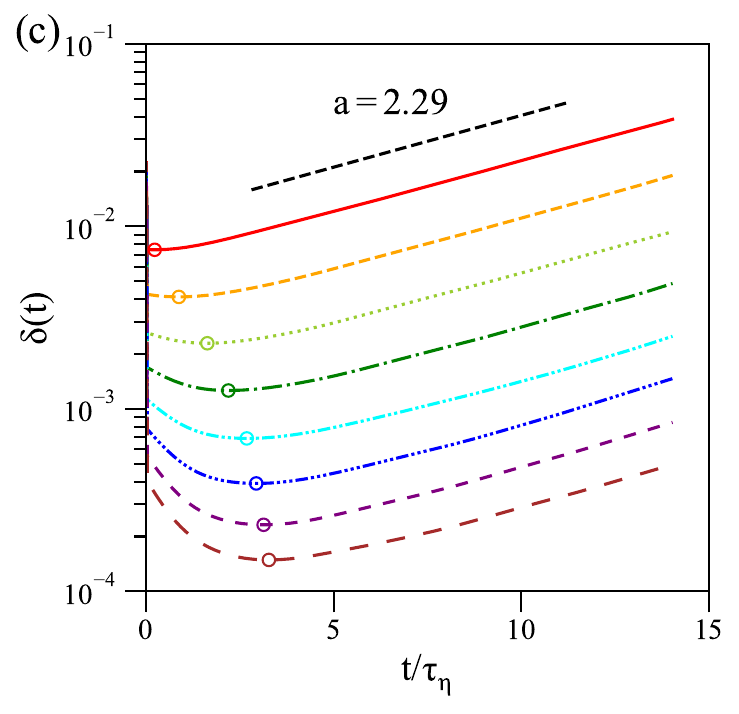}\hspace{-0.10in}
\caption{The temporal evolution of the error magnitude for the one-step data assimilation: (a) $N=64$, $Re_{\lambda}=60$; (b) $N=128$, $Re_{\lambda}=105$; (c) $N=256$, $Re_{\lambda}=160$.}\label{fig5}
\end{figure}

To explore the mechanism behind TSDA, it is natural to consider the one-step data assimilation (ODA), i.e. feeding in the reference data once in the start and observing the evolution of the error thereafter. This numerical experiment of ODA is illustrated in Fig.~\ref{fig5}, where the variations of the assimilation error are recorded with a range of assimilation wavenumbers for the cases listed in Table~\ref{tab1}. Again, four consecutive assimilation steps are performed initially, instead of one, to avoid the inconsistency caused by the initial Euler integration step used in the two-step Adams-Bashforth method.

As can be observed in Fig.~\ref{fig5}, the errors invariably experience an initial decaying period, followed by an exponential growth.The decay time before the error reaches the bottom is denoted by $t^*$. Apparently, this decrease of error is entirely due to the initial assimilation step. That is, a relaxation of the error occurs after the initial supply of the `true' data, making the error continuously decay for some time on the order of the Kolmogorov time scale. This relaxation process eventually ends with the error hitting the minimum as denoted by the circles in the figure. In this sense, the assimilation does not have to be continuous, and it would still be successful unless the time step is too large that the error has evolved far into the growing regime.

Here, a possible explanation for the behavior of the local error minimum in Fig.~\ref{fig5} is the competition between the two mechanisms: the decaying tendency of high-wavenumber error noise added at the initial time, and the growing tendency due to the chaotic nature of turbulence. At a larger $k_a$, more Fourier modes are enforced, the growth of errors should start from a lower initial error, resulting in a longer decay time $t^*$. The high-wavenumber error-relaxation mechanism will be further examined later in Section IIIC.

\begin{figure}\centering
\includegraphics[height=.38\textwidth]{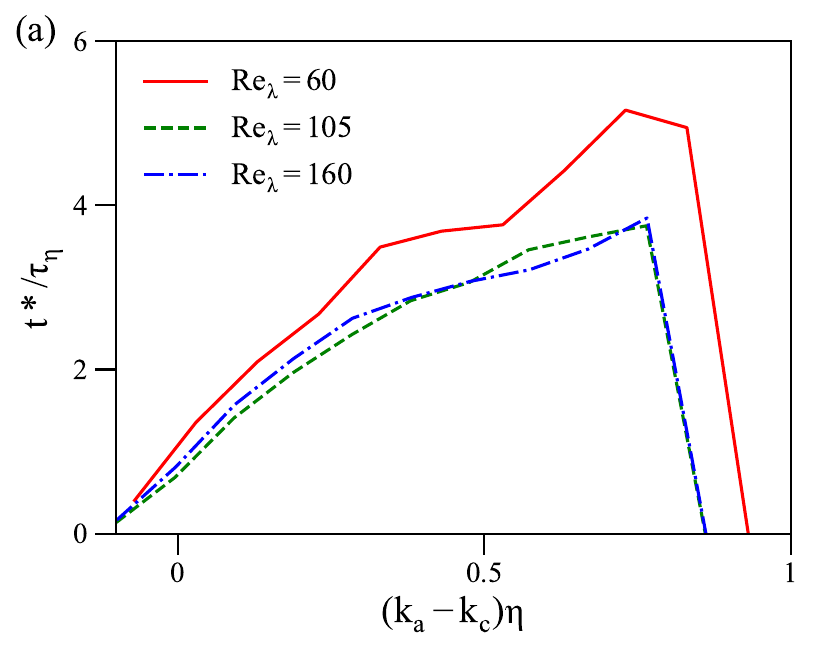}\hspace{-0.10in}\vspace{-0.10in}
\includegraphics[height=.38\textwidth]{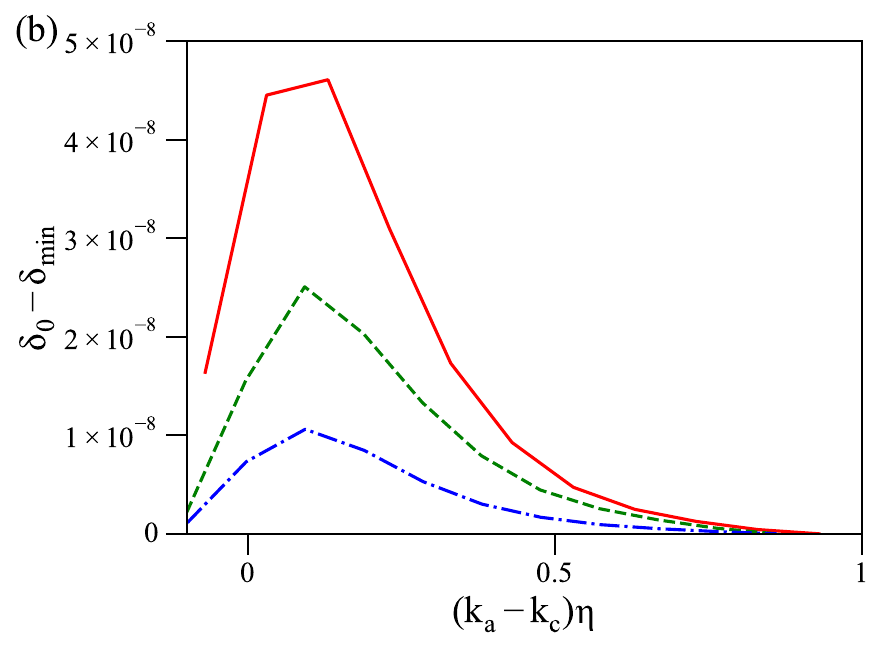}\hspace{-0.10in}\vspace{-0.10in}
 \caption{The decay time and the magnitude of the error decay in the one-step assimilation: (a) the decay time; (b) the magnitude of the error decay.}\label{fig6}
\end{figure}

The Kolmogorov-scale normalized decay time $t^* / \tau_{\eta}$ of ODA is plotted in Fig.~\ref{fig6}a against the scaled assimilation wavenumber $(k_a-k_c) \eta$. With this scaling, the three curves are close to each other. Due to the aforementioned mechanism, the error decay time increases with the assimilation wavenumber, but it eventually experiences a sharp drop to zero once the whole range of wavenumbers are used in ODA. In Fig.~\ref{fig6}b, the variation of the magnitude of the entire error decay, $\delta_0-\delta_{min}$, is plotted against the assimilation wavenumber. Here, $\delta_0$ represents the initial error and $\delta_{min}$ represents the minimum error shown in Fig.~\ref{fig5}. To further quantify the evolution of errors, we shall next examine both the initial decay rate and the subsequent growth rate of the error for ODA.

\begin{figure}\centering
\includegraphics[width=.5\textwidth]{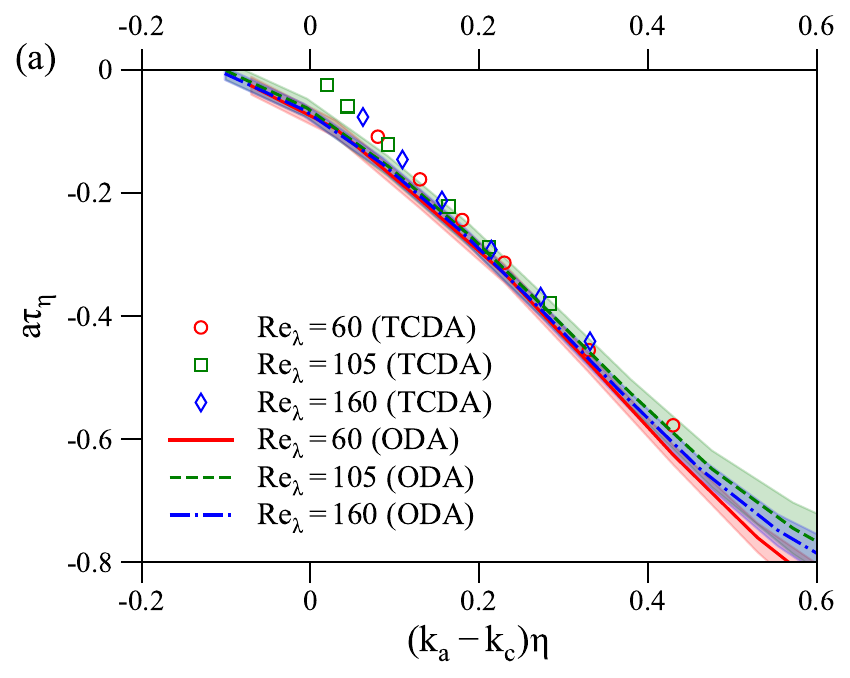}\hspace{-0.10in}
\includegraphics[width=.5\textwidth]{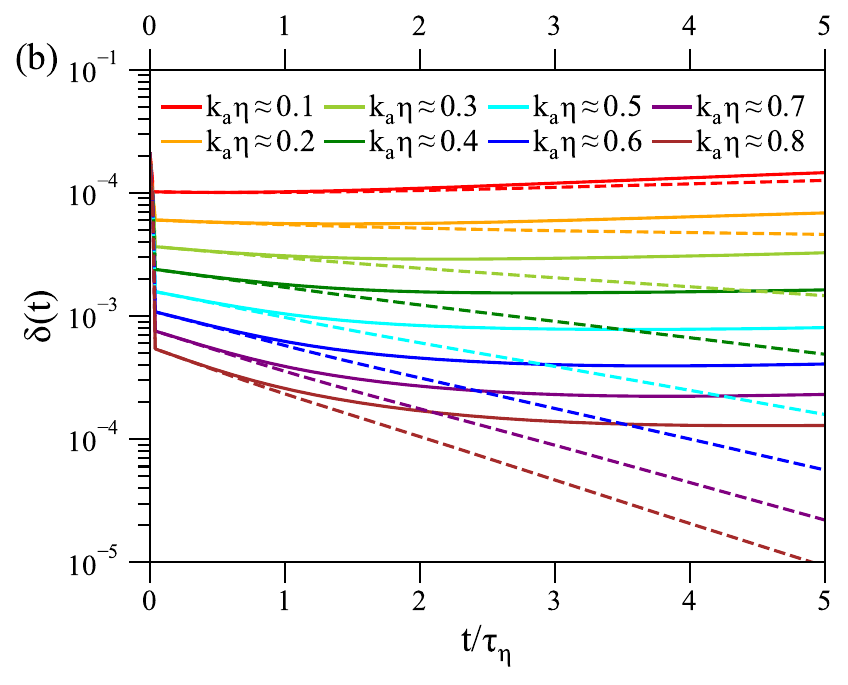}\hspace{-0.10in}
 \caption{The initial decay constant and the error evolution in the one-step assimilation: (a) decay constant, the shaded area represents the uncertainty when the initial slopes for the one-step assimilation are calculated using the first 1 to 30 time steps; (b) the initial evolution of errors with the solid line representing the one-step assimilation and the dashed line representing the continuous assimilation.}\label{fig7}
\end{figure}

The initial exponential decay constant $a$ for ODA is shown in Fig.~\ref{fig7}a, along with the decay constants for TCDA. Apparently, for ODA, the exponential behavior is only valid in the initial period. To account for the uncertainty in the calculation of the decay constant for ODA, 30 calculations are performed using the first 1, 2, 3 $\ldots$ 30 consecutive time steps. The shaded area in Fig.~\ref{fig7}a represents the range (or uncertainty) of the decay constants in ODA when the range of linear fitting is varied. As discovered in the previous work, \cite{Lalescu2013} the decay constants for TCDA follow closely a linear law as given by Eq.~(\ref{linearlaw}), with the data passing through the origin when scaled by $(k_a-k_c) \eta$. The slope constant $-\beta \approx -1.35$ as given in Section IIA. Interestingly, the initial decay constant for ODA is very close to that of TCDA. The exception is only for the very small assimilation wavenumbers. This should not be surprising since for a moderate level of error, the one-step assimilation is still expected to reduce the error for a short period even using a very small assimilation wavenumber. However, this could not occur if the initial error is relatively small, thus any assimilation cases (being continuous or not) with such assimilation numbers are eventually unsuccessful. A more direct view of the initial decay constant is shown in Fig.~\ref{fig7}b, where evolution of errors is shown for both TCDA and ODA. Clearly, ODA has the same error decaying rate as TCDA at the initial stage, but the decay rate gradually decreases with time.

\begin{figure}\centering
\includegraphics[width=.6\textwidth]{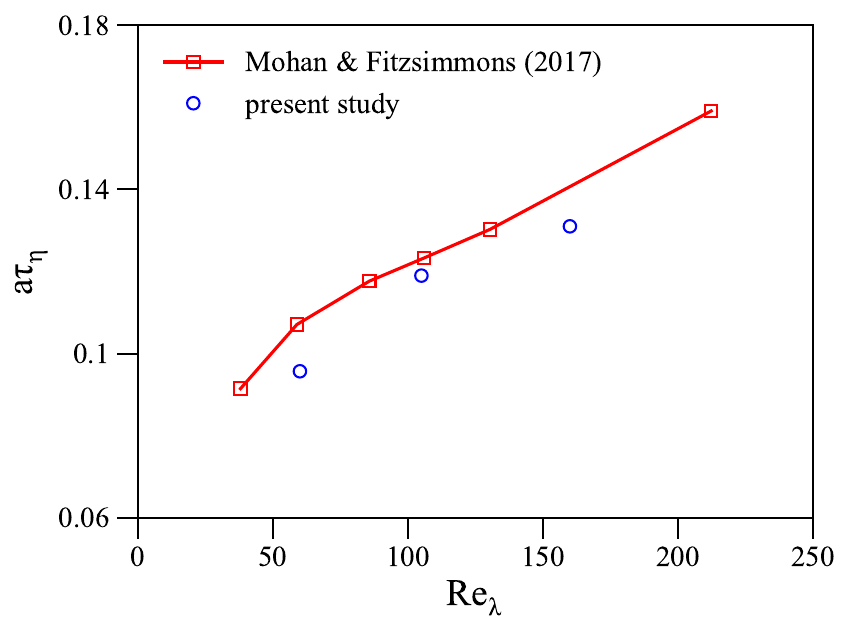}\hspace{-0.10in}
 \caption{The growth exponent with respect to the Taylor Reynolds number. Also shown in the figure is largest Lyapunov exponent curve reported by Mohan and Fitzsimmons. \cite{Mohan2017}}\label{fig8}
\end{figure}

Finally, it is interesting to examine the growing rate of errors after the ODA curves reach the minimum (cf. Fig.~\ref{fig5}), as it is related to the instability and chaotic behavior of turbulence. In this consideration, we plot the exponential growth constant against the Taylor Reynolds number in Fig.~\ref{fig8}. Also shown in the figure is the largest Lyapunov exponent curve for incompressible isotropic turbulence reported by Mohan and Fitzsimmons, \cite{Mohan2017} who have found that the largest Lyapunov exponent normalized by the Kolmogorov time scale increases with Taylor Reynolds number with saturation at large Reynolds number as apposed to stay universally constant. \cite{Crisanti1993,Aurell1996} As shown in Fig.~\ref{fig8}, this trend is also captured in the present study. It is well known in chaotic theory that a chaotic system would eventually be dominated by the largest Lyapunov exponent. \cite{Alligood1996} This is also demonstrated by the current analysis as we recall that, in Fig.~\ref{fig5}, after hitting their minimums, all the curves share a similar slope that depends on the Taylor Reynolds number, reflecting the dominance of the largest Lyapunov exponent.

From the discussions of Fig.~\ref{fig7}, it can be easily understood why TSDA can be as good as TCDA, since there is an initial period during which ODA has the same decaying rate as TCDA. However, it is yet unclear regarding the intriguing behavior that TSDA can have, in some cases, slightly larger decaying rate than TCDA (cf. Fig.~\ref{fig2} and Fig.~\ref{fig3}b). This issue will be addressed next.

\subsection{A Further analysis on the error spectrum in data assimilation}

To further scrutinize the evolution of errors in the data assimilation process, it is necessary to dissect the assimilation error into each length scale (wavenumber) so that the individual contribution from different length scales to the error decaying can be visualized. To this end, we define the error spectrum as \cite{Yoshida2005}

\begin{equation}
  E_{\delta}(k) = \sum_{\ k-\frac{1}{2} \le |\mathbf{k}| < k+\frac{1}{2}} \frac{1}{2}[\mathbf{\hat{u}}(\mathbf{k},t) - \mathbf{\hat{u}}^{ref}(\mathbf{k},t)]^2.
  \label{Ed}
\end{equation}
Apparently, the error spectrum should vanish everywhere at all wavenumbers if $\mathbf{u}$ converges to $\mathbf{u}^{ref}$.

\begin{figure}\centering
\includegraphics[width=.45\textwidth]{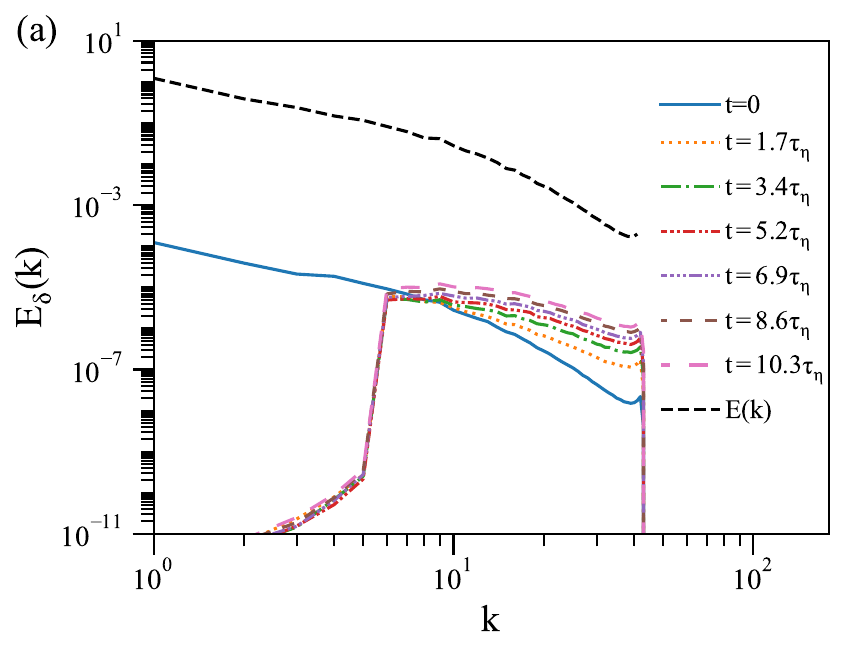}\hspace{-0.10in}
\includegraphics[width=.45\textwidth]{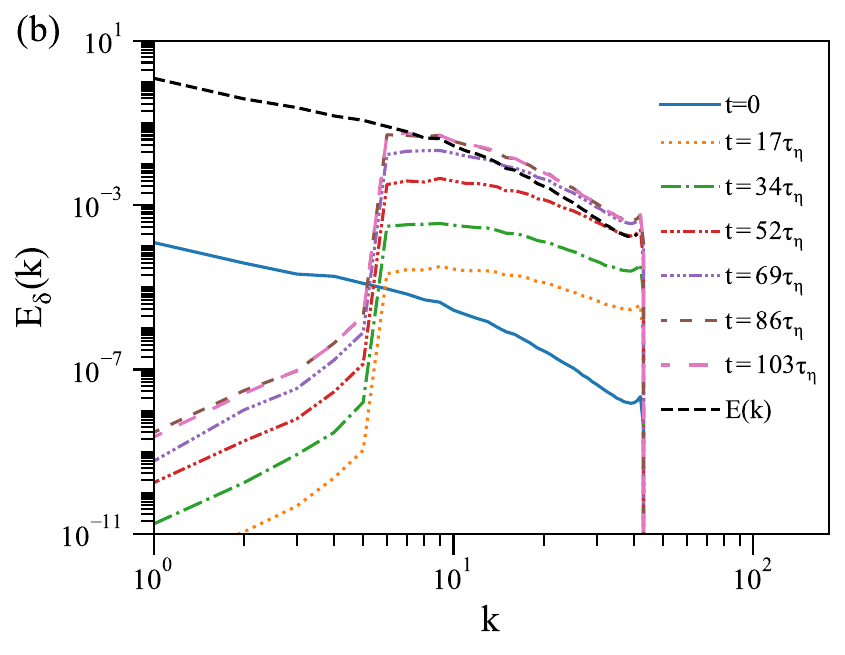}\vspace{-0.10in}

\includegraphics[width=.45\textwidth]{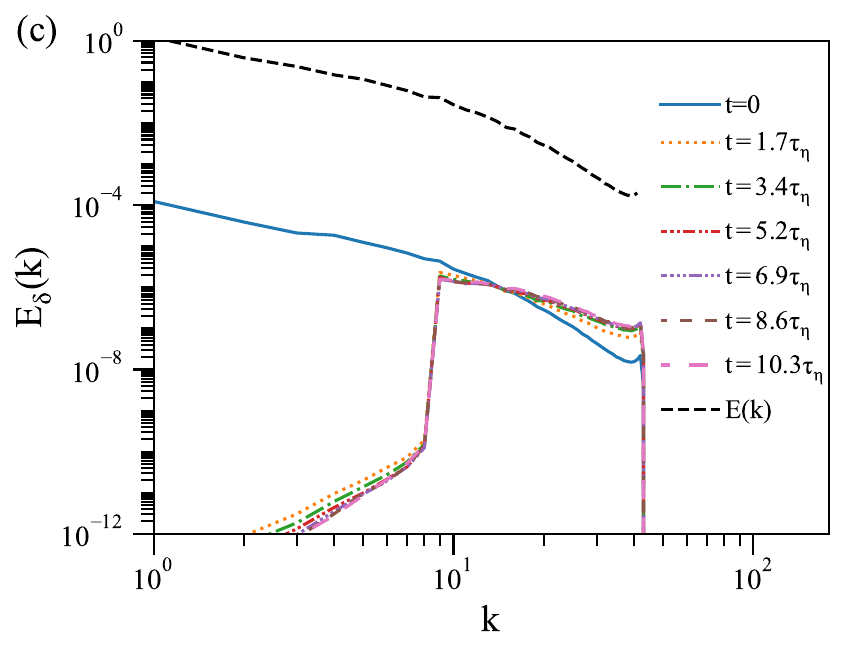}\hspace{-0.10in}
\includegraphics[width=.45\textwidth]{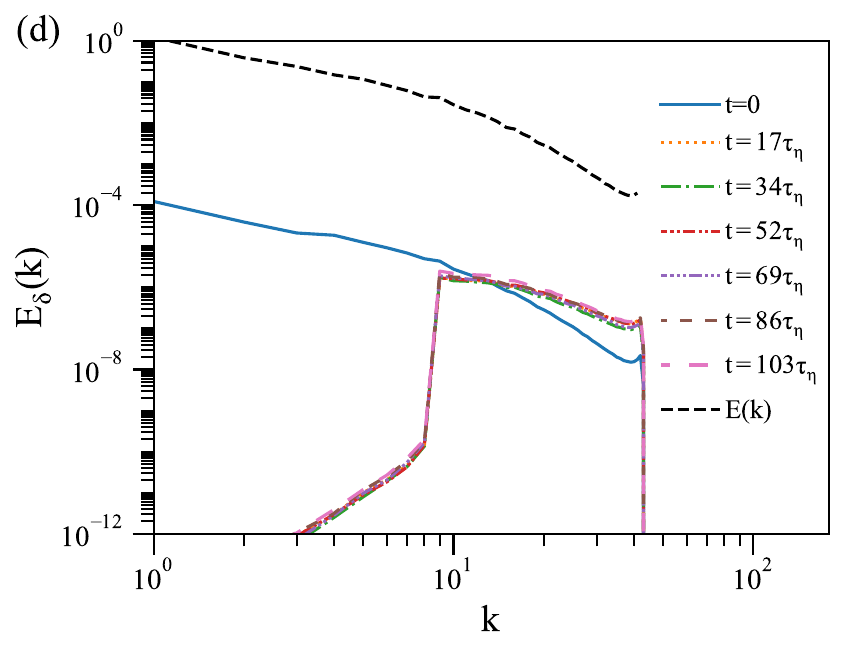}\vspace{-0.10in}

\includegraphics[width=.45\textwidth]{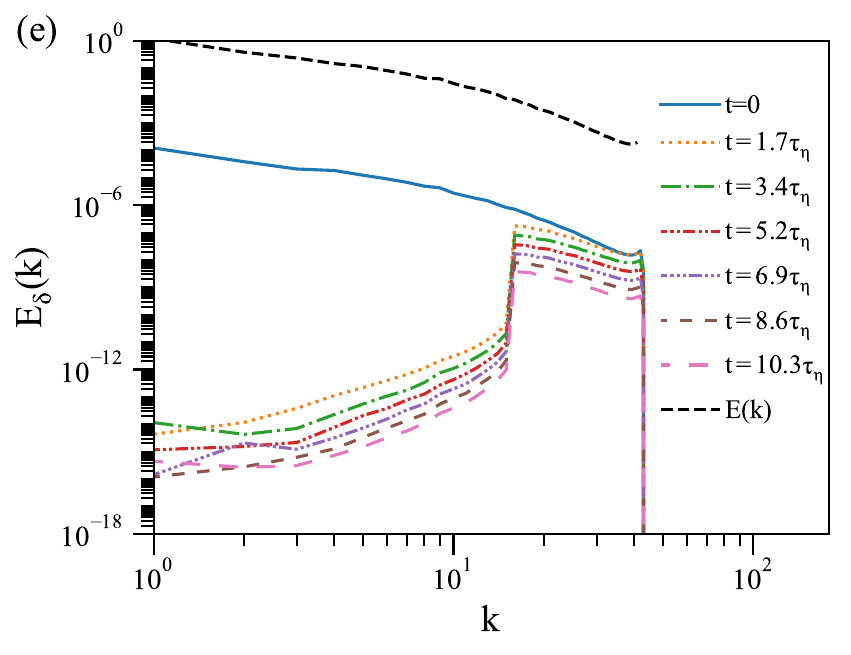}\hspace{-0.10in}
\includegraphics[width=.45\textwidth]{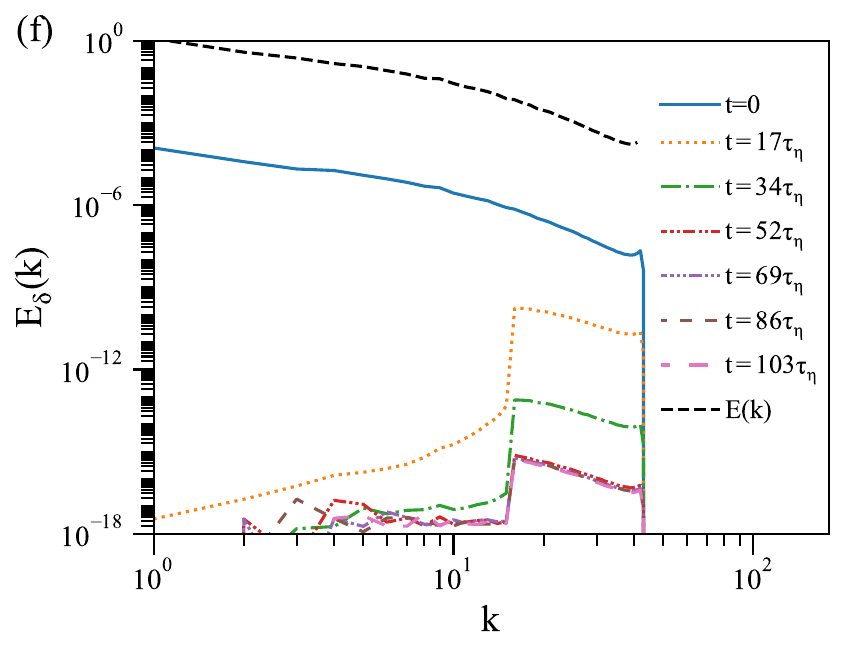}\vspace{-0.10in}
 \caption{The temporal evolution of the energy spectrum of the error field for the continuous data assimilation for the $Re_{\lambda}=105$ case: (a) $k_a \eta=0.12~(k_a=5),~0<t<10.3\tau_{\eta}$; (b) $k_a \eta=0.12~(k_a=5),~0<t<103\tau_{\eta}$; (c) $k_a \eta=0.19~(k_a=8),~0<t<10.3\tau_{\eta}$; (d) $k_a \eta=0.19~(k_a=8),~0<t<103\tau_{\eta}$; (e) $k_a \eta=0.36~(k_a=15),~0<t<10.3\tau_{\eta}$; (f) $k_a \eta=0.36~(k_a=15),~0<t<103\tau_{\eta}$.}\label{fig9}
\end{figure}

In Fig.~\ref{fig9}, the error spectrum for TCDA of the $Re_{\lambda}=105$ case is illustrated. The results for other cases are very similar and thus not reproduced. In the figure, three assimilation wavenumbers are considered, namely, $k_a \eta=0.12~(k_a=5)$ for the unsuccessful assimilation in Figs.~\ref{fig9}a and \ref{fig9}b, $k_a \eta=0.19~(k_a=8)$ for the neutral case (i.e. marginal assimilation) in Figs.~\ref{fig9}c and \ref{fig9}d, and $k_a \eta=0.36~(k_a=15)$ for the successful assimilation in Figs.~\ref{fig9}e and \ref{fig9}f. The correct energy spectrum $E(k)$ is also shown in all these figures for comparison. Here, we emphasize that the error spectrum is not calculated immediately after each data assimilation. Instead, the calculation is performed after the time integration, before the next round of large-scale data replacement in the subsequent time step, so that some small nonzero errors appear for $k<k_a$. Otherwise, the errors for $k<k_a$ should be exactly zero if the error spectrum is calculated immediately after the enforcement of the correct low-wavenumber data. As can be seen in the figure, when the assimilation wavenumber is insufficient (cf. Figs.~\ref{fig9}a and \ref{fig9}b), the errors invariably grow for the wavenumbers larger than $k_a$. Meanwhile, the errors at larger wavenumbers grow faster initially since small scales possess less energy and are thus more sensitive. Also, these scales are somehow less constrained since they are farther away from the assimilation wavenumber. Once the errors in the larger wavenumbers have grown to a certain level, the errors at wavenumbers near the assimilation wavenumber also start to grow due to the accumulation of errors at small scales. The growth of errors finally saturates at the same level of the energy spectrum of the true field, i.e. $E_{\delta}(k) \approx E(k)$ for $k>k_a$, indicating that $\mathbf{u}$ and $\mathbf{u}^{ref}$ are almost completely uncorrelated at high wavenumbers.

In the marginal case (cf. Figs.~\ref{fig9}c and \ref{fig9}d), there is initially a slight increase of error in the high wavenumber range and a weak decrease of errors for the wavenumbers close to the assimilation wavenumber. But this process soon comes to an end with the distribution of errors among all wavenumbers fixed and maintained at the same level, giving rise to the neutral state. Finally, for a successful data assimilation as shown in Figs.~\ref{fig9}e and \ref{fig9}f, the errors at all wavenumbers invariably decrease with time until saturating to the machine-error level with some osscilatory noises. After observing the error spectrum for TCDA, we are in a position to re-examine TSDA and make comparative analysis.  

\begin{figure}\centering
\includegraphics[width=.5\textwidth]{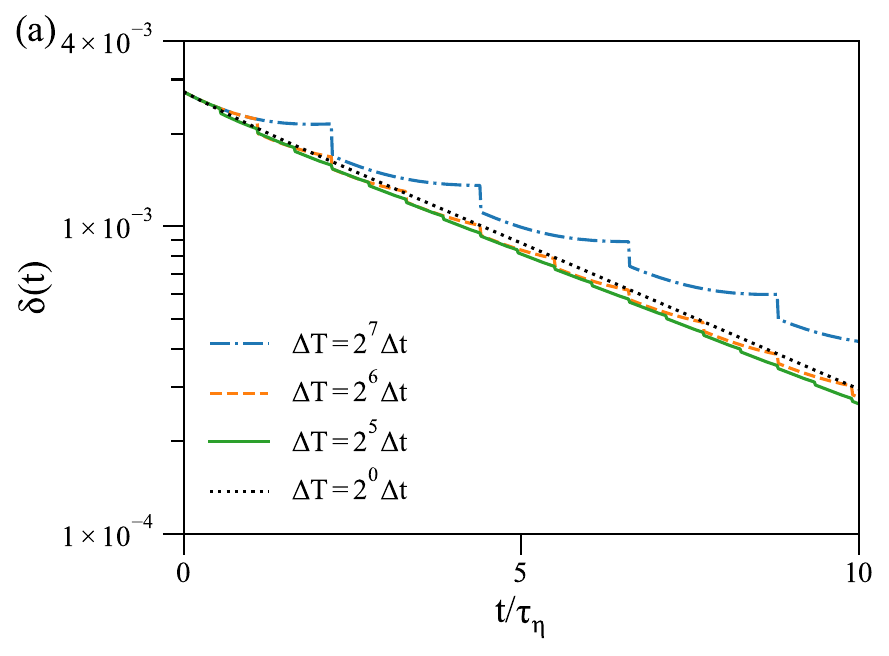}\hspace{-0.10in}
\includegraphics[width=.5\textwidth]{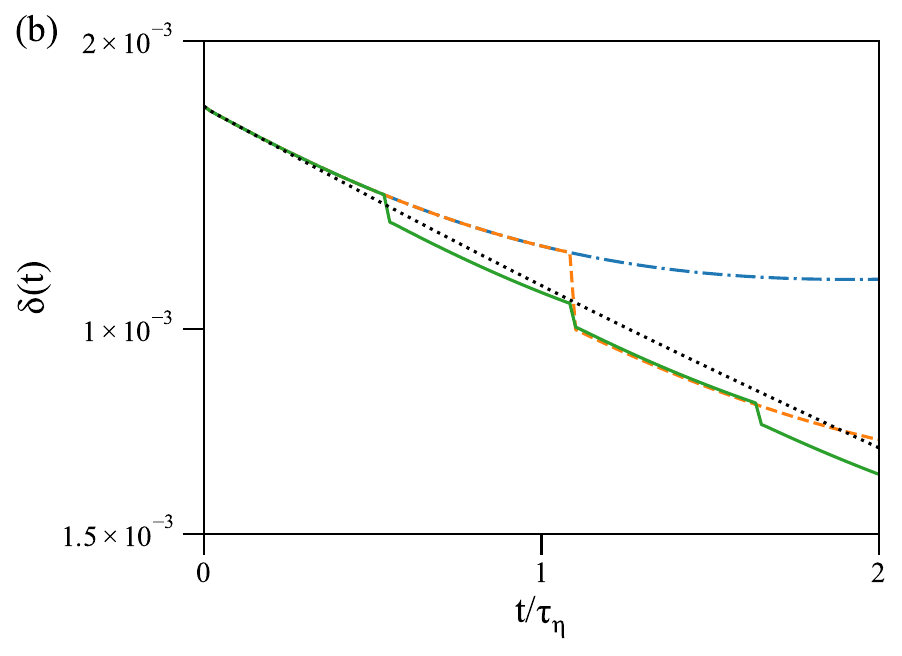}\hspace{-0.10in}
 \caption{The initial stage of the temporal evolution of the error magnitude for the temporally sparse and continuous data assimilation: (a) $0<t<10 \tau_{\eta}$; (b) $0<t<2 \tau_{\eta}$.}\label{fig10}
\end{figure}

To unravel why TSDA can have slightly larger decaying rate of errors, it is necessary to look closer into the error evolution at the very early stage of the assimilation. This is shown in Fig.~\ref{fig10}, where Fig.~\ref{fig10}b is a further zoom-in view of Fig.~\ref{fig10}a. Taking $\Delta T=2^5 \Delta t$ for instance, it is obvious that its initial decaying rate is the same as the continuous case (represented by the $\Delta T=2^0 \Delta t$ curve), but it gradually decreases with time. However, at each subsequent assimilation step, the error experiences a steep drop which may drag the error below and further below the continuous line such as in the $\Delta T=2^5 \Delta t$ case. It is precisely this interesting behavior that gives rise to the slightly faster error decaying rate of the large-step case over the continuous case. To further understand this behavior, it is necessary to compare the error spectra of the continuous and one-step data assimilation. Meanwhile, due to the fact that initial perturbation errors are contained at high wavenumbers, the dissipation spectrum of the error field, defined as $D_{\delta}(k)=2\nu k^2 E_{\delta}(k)$,\cite{Pope2000} will also be discussed in the following. 

\begin{figure}\centering
\includegraphics[width=.34\textwidth]{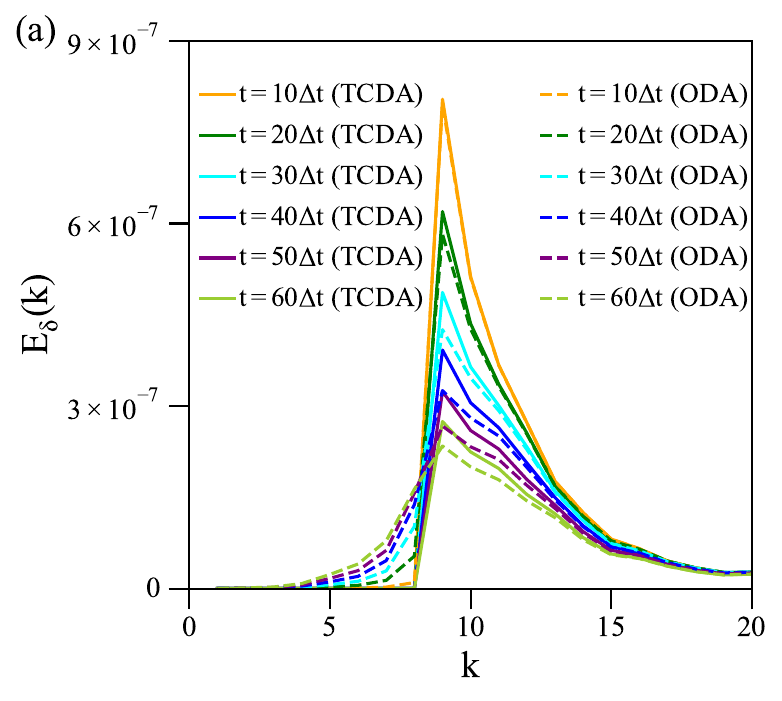}\hspace{-0.10in}
\includegraphics[width=.34\textwidth]{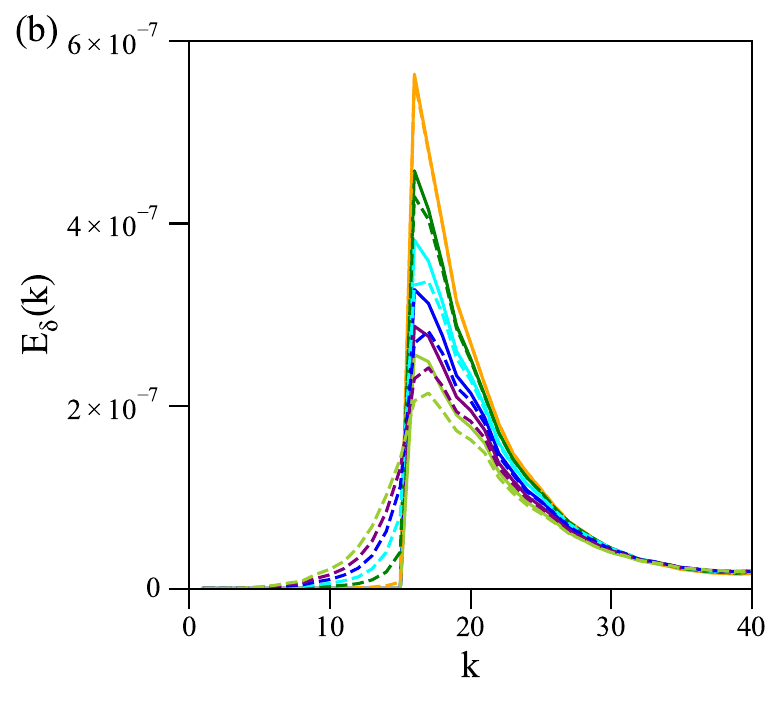}\hspace{-0.10in}
\includegraphics[width=.34\textwidth]{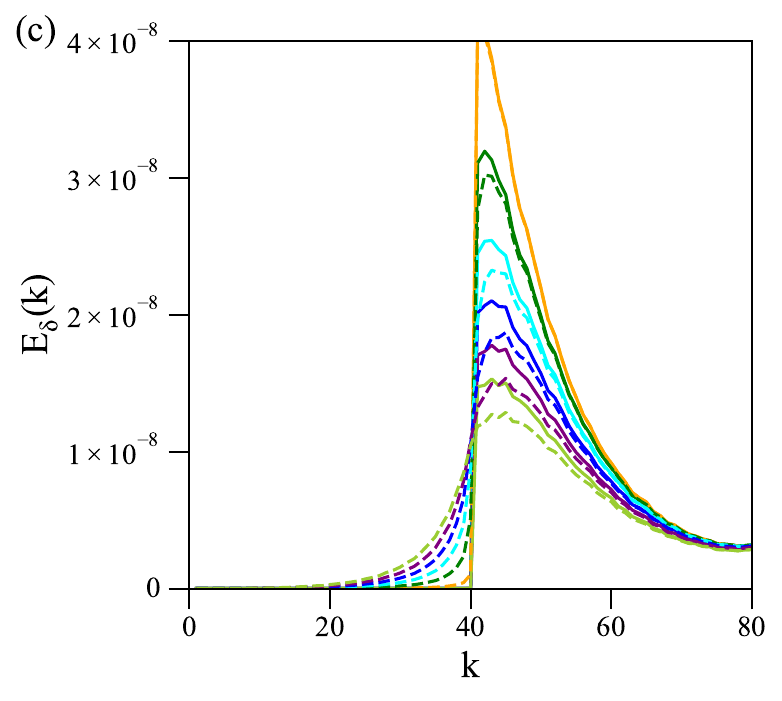}\hspace{-0.10in}

\includegraphics[width=.34\textwidth]{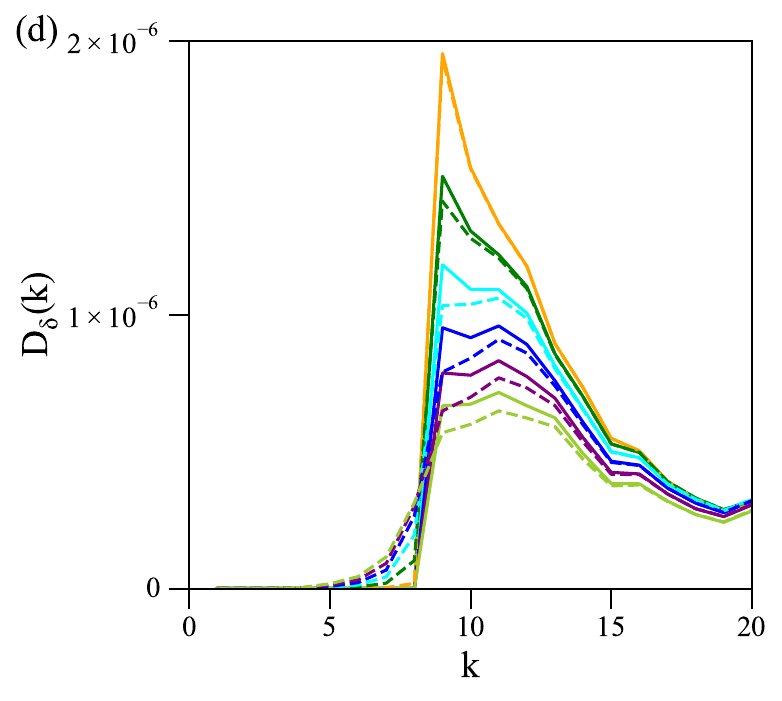}\hspace{-0.10in}
\includegraphics[width=.34\textwidth]{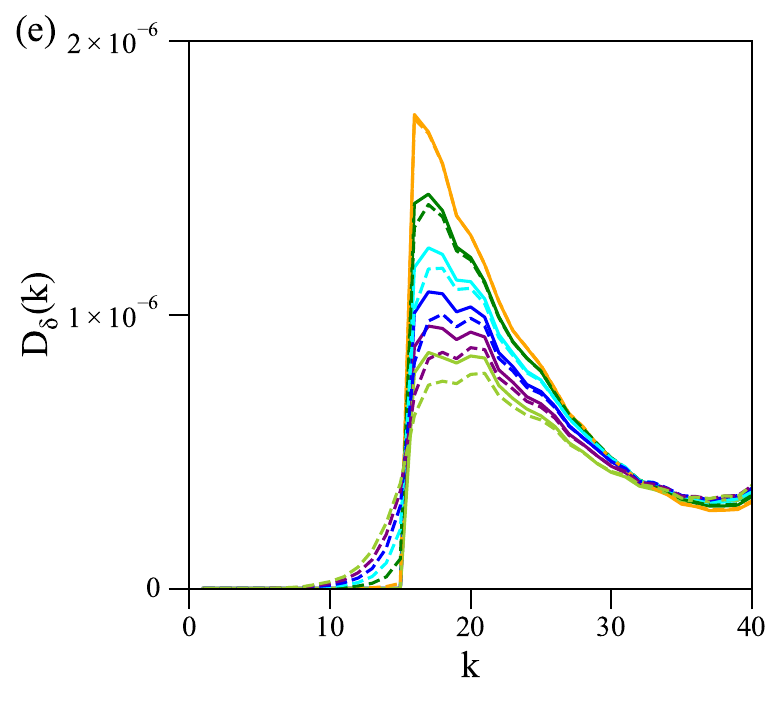}\hspace{-0.10in}
\includegraphics[width=.34\textwidth]{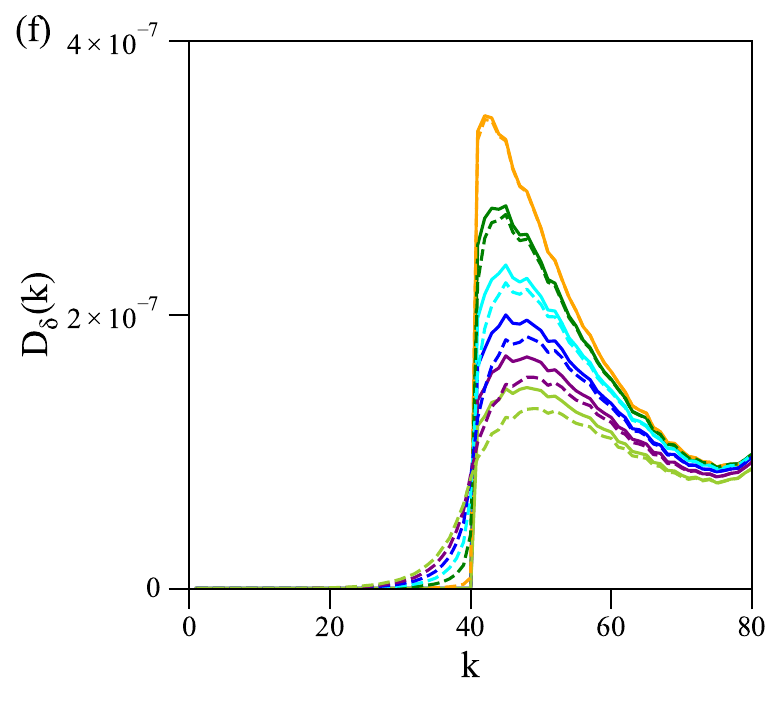}\hspace{-0.10in}
 \caption{The temporal evolutions of the energy spectrum $E_{\delta}(k)$ and the dissipation spectrum $D_{\delta}(k)$ of the assimilation error field in the both the one-step and the continuous cases: (a) $E_{\delta}(k)$, $N=64$, $Re_{\lambda}=60$; (b) $E_{\delta}(k)$, $N=128$, $Re_{\lambda}=105$; (c) $E_{\delta}(k)$, $N=256$, $Re_{\lambda}=160$; (d) $D_{\delta}(k)$, $N=64$, $Re_{\lambda}=60$; (e) $D_{\delta}(k)$, $N=128$, $Re_{\lambda}=105$; (f) $D_{\delta}(k)$, $N=256$, $Re_{\lambda}=160$.}\label{fig11}
\end{figure}

The energy spectrum and dissipation spectrum of the error field for TCDA and ODA are shown in Fig.~\ref{fig11} for all three cases listed in Table~\ref{tab1}. Here the assimilation wavenumber $k_a=8,\ 15$ and $40$ for the $Re_{\lambda}=60,\ 105$ and $160$ cases, respectively. Also, the normal coordinate scale is used, instead of the widely adopted log-log scale, to more directly visualize the magnitudes of the energy spectrum and dissipation spectrum of the error field from each wavenumbers.

From Figs.~\ref{fig11}a to c, one immediately observes that the errors at $k>k_a$ for ODA drop faster than that for TCDA, indicating that, given some free-relaxation time, the errors at small scales can drop faster than the continuously enforced case. In turn, ODA also has less burden to dissipate the error as reflected in the dissipation spectrum. In fact, the differences between dissipation spectra for ODA and TCDA are more pronounced considering the larger ranges in the vertical coordinate axis with Figs.~\ref{fig11}d to f, compared to that with Figs.~\ref{fig11}a to c. One can also observe that, while the errors above the assimilation wavenumber decrease with time, the errors below the assimilation wavenumber increase since no large-scale data is available after the initial supply. On the other hand, the errors at all scales decrease with time in TCDA since the assimilation is performed continuously. At the beginning, the advantage of ODA above the assimilation wavenumber is roughly in balance with its disadvantage below the assimilation wavenumber compared to TCDA, such that both ODA and TCDA share a similar decaying rate initially. Later, the difference between ODA and TCDA becomes larger as the large-scale errors grow too much for the former.

However, this would not happen in TSDA since new data in future time steps comes in and annihilates the large-scale deficit for ODA while its small scale advantage is maintained. This process is exactly what behind the slightly faster decaying rate for some TSDA cases, and is directly responsible for the sharp drop of total errors as shown in Fig.~\ref{fig10}. Obviously, the time step cannot be too large, otherwise the deficit would be too much to overcome by the next supply of data. Also, TSDA only slightly outperforms TCDA in terms of error dropping rate, since once the error drops, its advantage at larger wavenumbers would also drop. Nevertheless, its much-reduced amount of required data and consequently the higher computational efficiency are definitely meaningful.

At this point, though the mechanism of TSDA is to some extent unraveled, the mathematical or physical reasons regarding why the errors relax faster for TSDA than TCDA are still not clear. We conjecture that the errors at machine level is potentially a cause to this phenomenon. It is understood that, even though one can deem a particular DNS solution as the `correct' solution, this solution cannot be exactly stored unless with a machine with infinite accuracy. In this sense, the true solution $\mathbf{u}^{ref}$ inevitably contains some small errors. These round-off errors would affect the assimilation process such that the error eventually fluctuates around the machine-error level, instead of going exactly to zero.

\begin{figure}\centering
\includegraphics[width=.5\textwidth]{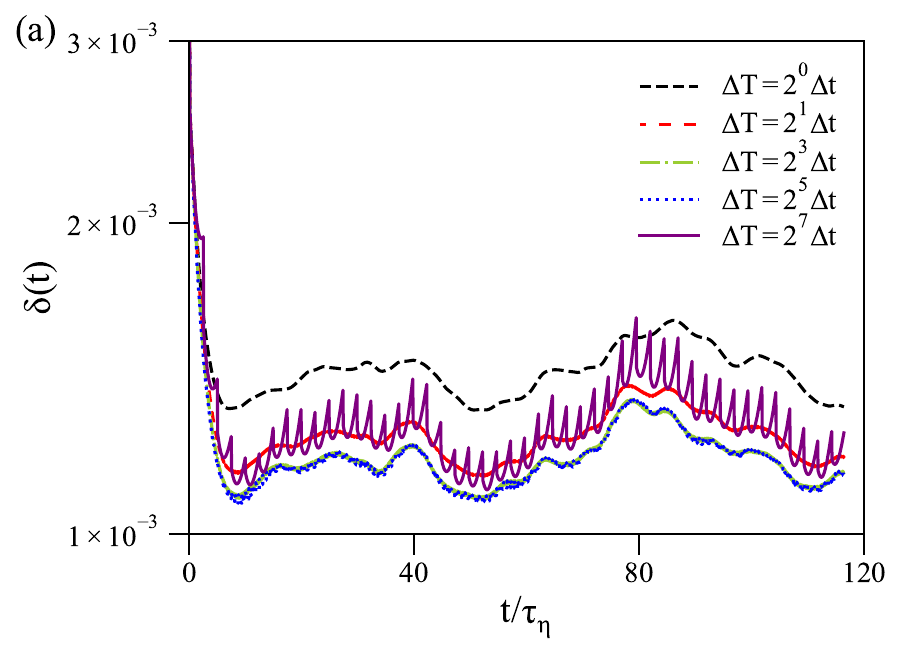}\hspace{-0.10in}
\includegraphics[width=.5\textwidth]{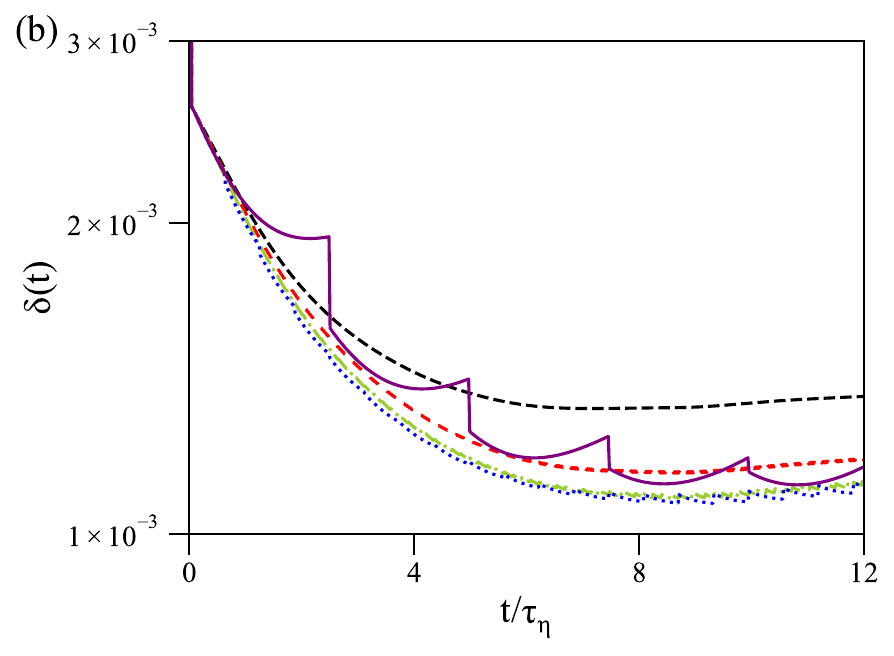}\vspace{-0.10in}
 \caption{The temporal evolution of the error magnitude of the data assimilation using a reference field with incorporated non-negligible large-scale errors for the $N=64$ case: (a) $0<t<120 \tau_{\eta}$; (b) $0<t<12 \tau_{\eta}$.}\label{fig12}
\end{figure}

As a result, TCDA introduces more errors (albeit small) than TSDA. While these errors will not affect the enforced large scales, they unavoidably propagate into small scales due to the inherent cascade property of turbulence, and consequently, overload the high-wavenumber error-relaxation process discussed in Section IIIB. Even so, the difference of the final error between TSDA and TCDA is invisible since the machine-error is relatively too small and also contains randomness. To more clearly confirm the effect of errors in the large-scale flow fields, we artificially contaminate $\mathbf{u}^{ref}$ by incorporating errors much larger than the machine round-off level. The results for the evolution of errors are shown in Fig.~\ref{fig12} for the $N=64$ case for demonstration. Here, the reference field is replaced by an erroneous field, given as
\begin{equation}
  \mathbf{\hat u}^{e}(\mathbf{k},t)=(1+\varepsilon_e)\mathbf{\hat u}^{ref}(\mathbf{k},t),
  \label{ue}
\end{equation}
where $\varepsilon_e=1\times 10^{-3}$ is deliberately chosen such that $\varepsilon_e<\varepsilon$ (we recall that $\varepsilon$ is the perturbation in the initial condition of $u$). In this case, it is still reasonable to expect that the magnitude of error $\delta$ should drop to the order of $10^{-3}$ if the assimilation is successful. In Fig.~\ref{fig12}, the results for both TCDA and TSDA with different time steps are included. Indeed, a drop of errors to the level of $10^{-3}$ occurs for the tested cases. As expected, at the end of the assimilation, TCDA has an error about 35 percent larger than TSDA with a properly chosen time step. In the zoom-in view as shown in Fig.~\ref{fig12}b, one also clearly recognizes the faster decaying rate of error for the large-step case (see the $\Delta T=2^5 \Delta t$ case for instance).

\begin{figure}\centering
\includegraphics[width=.6\textwidth]{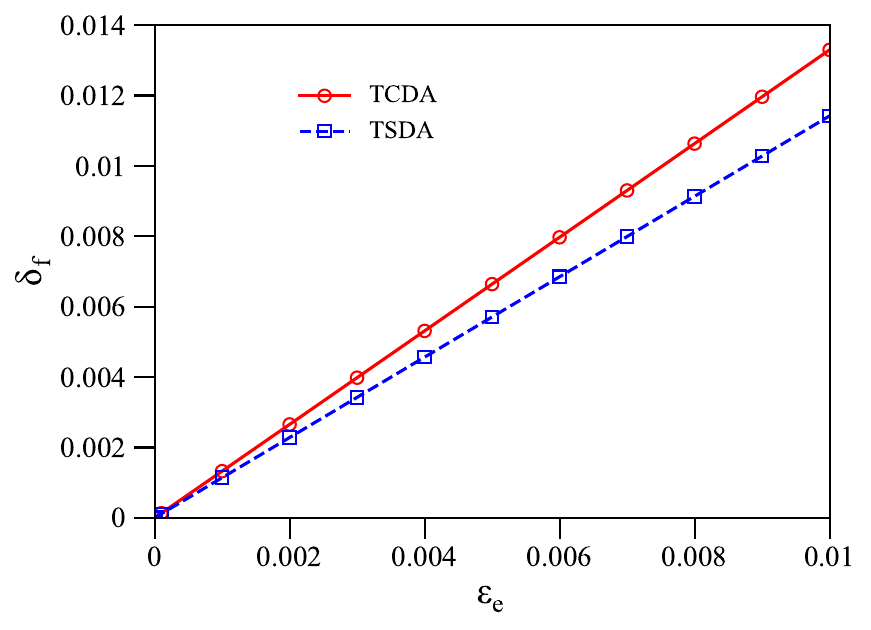}\hspace{-0.10in}
 \caption{The final error magnitude for the temporally continuously and sparse data assimilation with incorporated non-negligible large-scale errors for the $N=64$ case.}\label{fig13}
\end{figure}

A more straightforward visualization of the difference between the performance of TCDA and TSDA is given in Fig.~\ref{fig13} where the artificial error ranges from $\varepsilon_e=1\times 10^{-3}$ to $10^{-2}$. Here again, the displayed result is for the $N=64$ case, and $\Delta T=32 \Delta t$ is chosen for TSDA. As we can see, for both TCDA and TSDA, the errors at the end of the assimilation linearly increase with the large-scale errors. Meanwhile, TCDA clearly ends up with more errors than TSDA as the errors in the large scales increase. Here, using the least square fitting, the roughly linear behavior of the final assimilation error can be written as
\begin{equation}
  \delta_f=c \varepsilon_e,
  \label{errorwithepsilon}
\end{equation}
where the constant $c=1.33$ for TCDA and $1.14$ for TSDA. As a consequence, it is potentially more beneficial to adopt TSDA if the reference data contains non-negligible errors.

\section{The case of higher Reynolds number}

Up to this point, the presented results are for $Re_{\lambda}$ ranging from 60 to 160 as given in Table~\ref{tab1}, corresponding to the investigated values by Yoshida \emph{et al.} \cite{Yoshida2005} In this section, we further show some results for a higher Reynolds number at $Re_{\lambda} \approx 280$, which is higher than both the values investigated by Yoshida \emph{et al.} \cite{Yoshida2005} and those by Lalescu \emph{et al}.\cite{Lalescu2013} The numerical parameters are shown in Table~\ref{tab2}. In this case, some slightly different behaviors are observed for both TCDA and TSDA.

\begin{table}
\begin{center}
\small
\begin{tabular*}{0.8\textwidth}{@{\extracolsep{\fill}}ccccccc}
\hline
Reso. &$Re_{\lambda}$ &$\nu$ &$\Delta t$ &$\epsilon$ &$\eta$ &$k_{max} \eta$ \\ \hline
$512^{3}$  &280 &0.001 &0.0004 &0.69 &0.0062 &1.06\\ \hline
\end{tabular*}
\normalsize
\caption{The key numerical parameters for $Re_{\lambda}=280$.}\label{tab2}
\end{center}
\end{table}

\begin{figure}\centering
\includegraphics[width=.7\textwidth]{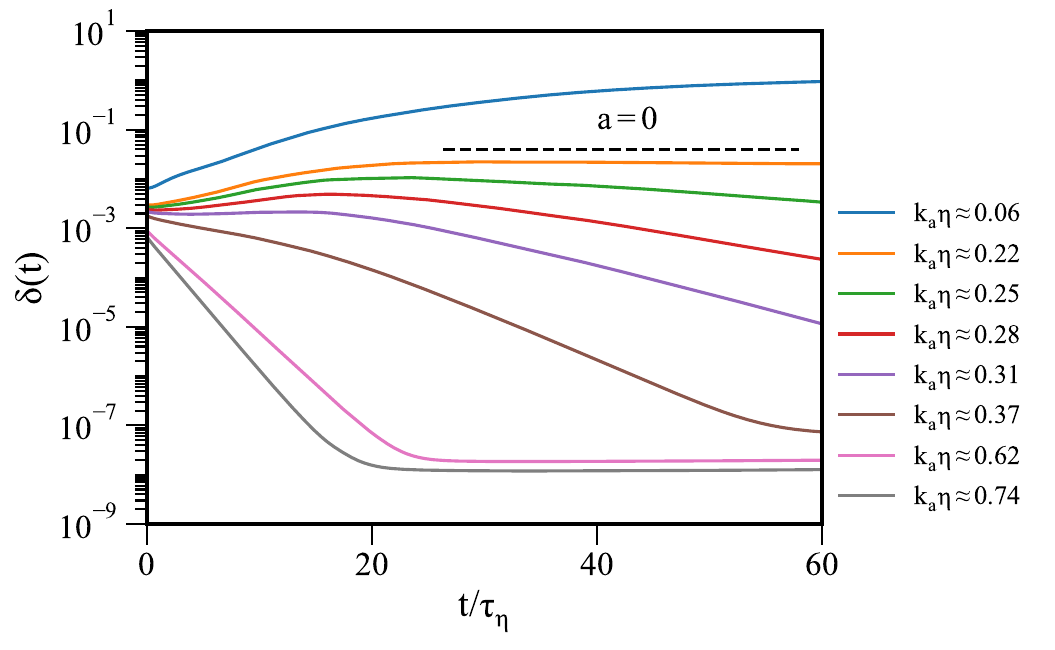}\hspace{-0.10in}
 \caption{The error magnitude evolution in TCDA at $Re_{\lambda} = 280$.}\label{fig14}
\end{figure}

The results for TCDA is shown in Fig.~\ref{fig14}. As observed, some slightly different behaviors are present in that there is no well-identified neutral state. Among all curves, $k_a \eta=0.22$ can be most likely defined as `neutral', even though the error first increases and then saturates to a plateau. It is worth emphasizing that this saturation is not the same as the saturation when the errors at $k>k_a$ grows uncontrollably and saturates due to the final balance between the dissipation and the applied large-scale forcing. This is evident since the magnitude of the unassimilated modes is well above $10^{-1}$, while the saturated error for $k_a \eta=0.22$ is only about $2\times 10^{-2}$. In other words, if the supply of the large-scale reference data is stopped in the saturation region, the errors at $k>k_a$ would continue to grow.

Even with this definition of the neutral curve, we still see that the critical wavenumber $k_c$ slightly increases with $Re_{\lambda}$, i.e. $k_c \eta \approx 0.175$ for $Re_{\lambda}=60$, $k_c \eta \approx 0.19$ for $Re_{\lambda}=105$, $k_c \eta \approx 0.21$ for $Re_{\lambda}=160$ (cf. Fig.~\ref{fig1}), and $k_c \eta \approx 0.22$ for $Re_{\lambda}=280$. This trend is also observed by Lalescu \emph{et al.}, \cite{Lalescu2013} who attribute this to intermittency.

\begin{figure}\centering
\includegraphics[width=.49\textwidth]{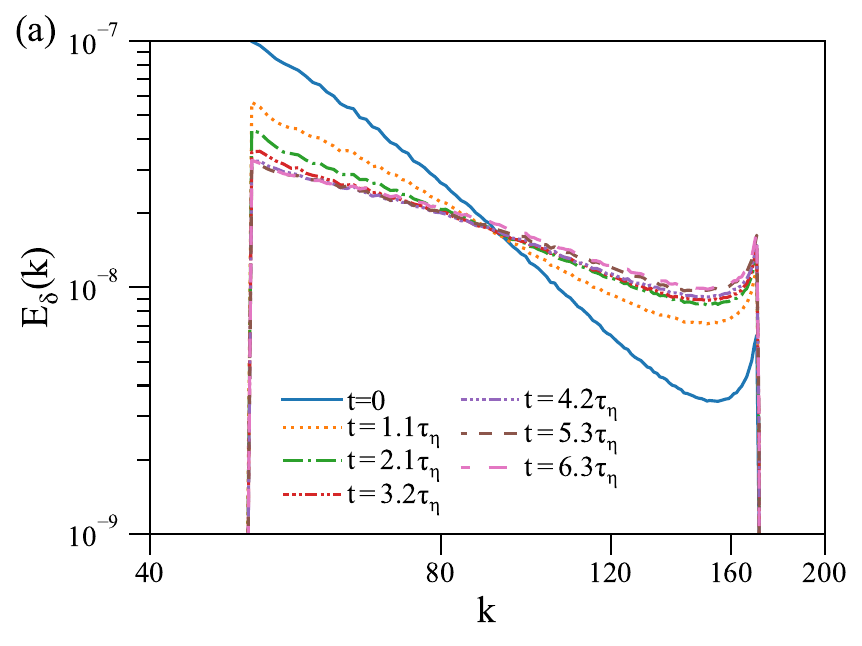}\vspace{-0.10in}
\includegraphics[width=.49\textwidth]{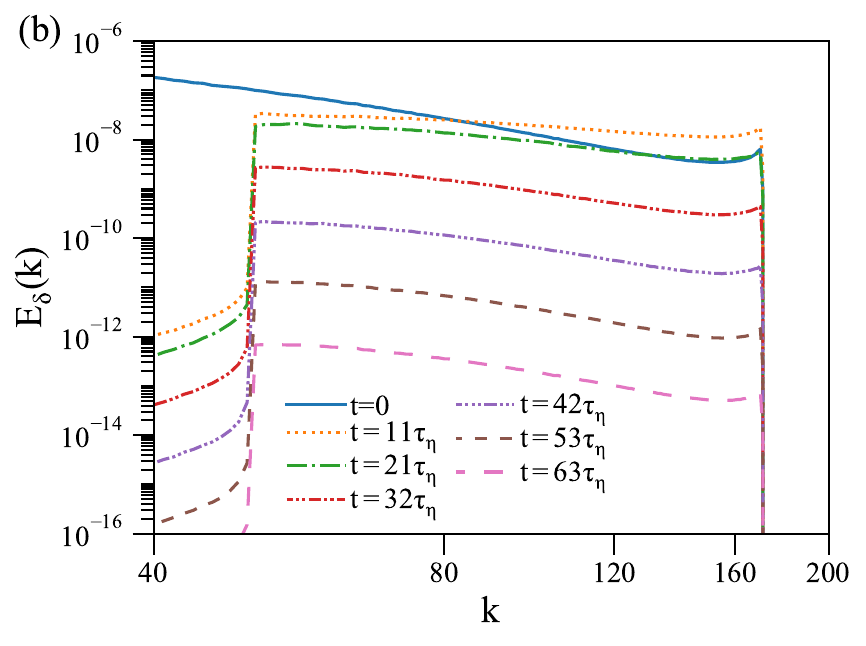}\vspace{-0.10in}
 \caption{The temporal evolution of the energy spectrum of the error field for TCDA at $Re_{\lambda} = 280$ and $k_a \eta=0.31$: (a) $0<t<6.3\tau_{\eta}$; (b) $0<t<63\tau_{\eta}$.}\label{fig15}
\end{figure}

Another interesting observation in Fig.~\ref{fig14} is that, some of the successful assimilation cases (e.g. $k_a \eta=0.28$ and $0.31$) experience a transition period before the exponential decay of errors, and the errors may also increase initially during the transition. To understand this, we show the error spectrum defined by Eq.~(\ref{Ed}) in Fig.~\ref{fig15} for $k_a \eta=0.31~(k_a=50)$. It is interesting to observe that, in the transition period (Fig.~\ref{fig15}a), the errors at $k>k_a$ behave differently among different modes: the errors in the relatively lower wavenumbers decrease while those in the higher wavenumbers increase. Only in the exponential error decay period do the errors invariably decrease among all wavenumbers (Fig.~\ref{fig15}b). This is different from the cases for relatively lower $Re_{\lambda}$ (cf. Figs.~\ref{fig9}e and \ref{fig9}f), where the errors at all wavenumbers decay throughout the assimilation period. We infer that this can be a result of the following mechanism: at a higher Reynolds number, the energy containing range is more separated from the dissipation range. In this case, for a $k_a$ that is sufficient for the small-scale reconstruction but relatively small, the effects of the enforced low-wavenumber modes cannot fully reach the dissipation range immediately. Instead, only the Fourier modes close to the low-wavenumber range (i.e. the intermediate range) most strongly feel the enforced modes initially. The errors in the far dissipation range only start to drop after the errors in the intermediate range have sufficiently decreased, giving rise to the exponential error decaying region.

\begin{figure}\centering
\includegraphics[width=.49\textwidth]{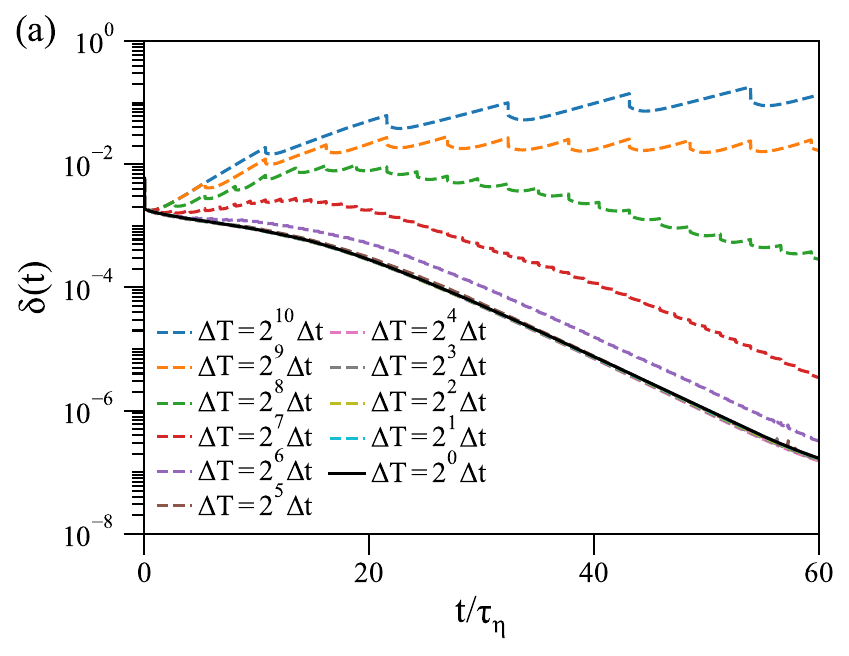}\vspace{-0.10in}
\includegraphics[width=.49\textwidth]{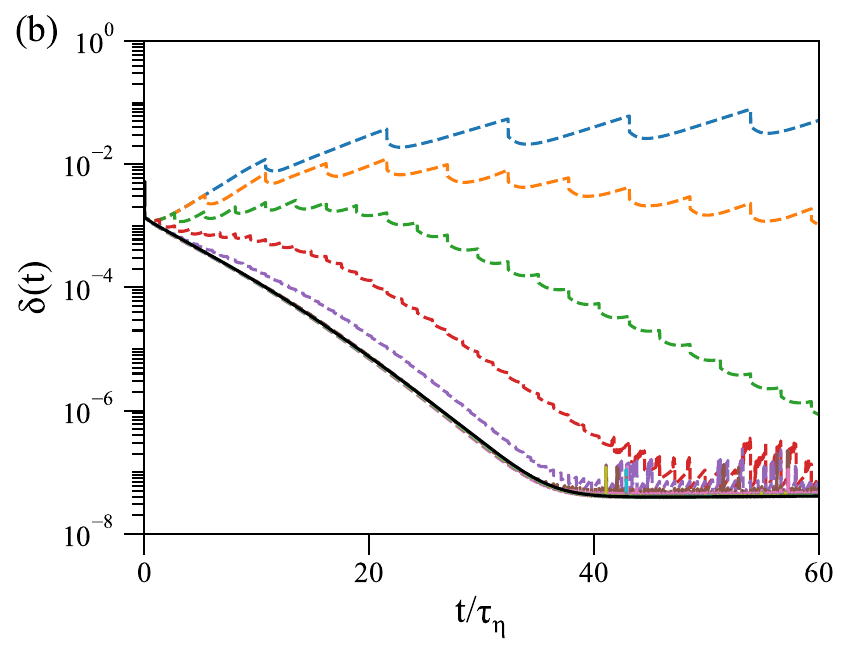}\vspace{-0.10in} \caption{The error magnitude evolution in TSDA at $Re_{\lambda} = 280$: (a) $k_a \eta=0.35$; (b) $k_a \eta=0.45$.}\label{fig16}
\end{figure}

Finally, we show the TSDA results in Figs.~\ref{fig16}a and \ref{fig16}b for $k_a \eta=0.35$ and $0.45$, respectively. As observed, at the current Reynolds number ($Re_{\lambda}=280$), for an assimilation condition which is sufficient for the small-scale reconstruction using TCDA, the TSDA can achieve a similar performance while $\Delta T$ is at least one order larger than $\Delta t$. The results for further increasing the Reynolds number will be pursued in future works.

%%%%%%%%%%%%%%%%%%%%%%%%%%%%%%%%%%%%%%%%%%%%%%%%%%%%%%%%%%%%%%%%%%%%%%%%%%%%%%%%%%%%%%%%%%%%%%%%%%%%%%%%%%%%%%%%%%%%%%%%%
\section{Investigation on the possibility of the sub-Kolmogorov-scale recovery using data assimilation}

As reported by several previous works, \cite{Anselmet1984,Paladin1987,Yakhot2005,Schumacher2007,Gibbon2012} flow structures smaller than the Kolmogorov length scales exist due to the spatial intermittency and cannot be resolved by the Kolmogorov scale resolution that is generally adopted in DNS. On the other hand, since the small scales are slaved to the large scales based on the DA experiment, it is tempting to think that the small scales should be recoverable by the DNS of Kolmogorov scale resolution. However, the question is more intriguing than it seems due to the intrinsic issue of the grid resolution for a highly nonlinear problem like the NS system. It should be emphasized that, in the DA scheme of previous sections as well as those in previous works, \cite{Yoshida2005,Lalescu2013} the reference data comes from the same grid resolution as the DA-based numerical solution. In other words, the large-scale data are obtained by projecting the fine solution to a coarser Fourier grid. In this case, to recover the sub-Kolmogorov scales, one would have to use finer grids so as to `recover' them, as if they are computed from the fine grid that is `sufficient' to capture all sub-Kolmogorov scales. With the DNS from a Kolmogorov resolution, however, the reference data is not from a projection of the true fine-grid solution, but from a coarser grid solution instead (i.e. the Kolmogorov resolution). This question is specifically raised in Ref \citenum{Lalescu2013}.

Indeed, numerical solutions with coarser grid is not expected to give the same result as that with a finer grid for highly nonlinear problems such as the NS equation. This implies that the large-scale data obtained from the DNS of Kolmogorov resolution is not the same as that obtained through a finer grid DNS, and the discrepancy is expected to grow with time. Nevertheless, it is fundamentally interesting to examine what exactly would happen if DA is applied in such cases.

For such a test, we first need a DNS of a sufficient resolution such that the sub-Kolmogorov scales can also be resolved and represented. In the current study, we choose the grid resolution $N=128$ for this `correct' reference simulation with the grid spacing $h_{DNS}=2 \pi/N$ slightly below $\eta /2$. Consequently, at least the scales at $\eta /2$ can be represented, and we assume that to be all the sub-Kolmogorov scales. Meanwhile, we run a second simulation with grid $N=64$ such that $h_{DNS} \approx \eta$, which is taken as a sufficient Kolmogorov-resolution. The key numerical parameters are given in Table~\ref{tab3}. Since the simulation case with the smallest grid spacing (i.e. $N=128$ in the current section) should at least satisfy $\eta/h_{DNS}>2$ for the presence of the sub-Kolmogorov scales, a smaller $Re_{\lambda}$ is chosen to lower the computational cost. The initial conditions for both cases are identically taken from the large scales ($k\le 10$) of a fully-developed turbulent field $\mathbf{u_0}$. In addition, the small scales are invariably set to zero. Consequently, we have the following initial conditions for both cases

\begin{equation}
\mathbf{\hat u}(\mathbf{k},t_0)=\binom{\mathbf{\hat u}_0(\mathbf{k}),\ \text{if}\ k \le 10}{0,~~~~~~~ \text{if}\ k >10},
\label{sub_Kolmogorov}
\end{equation}
where $k\le 10$ is intentionally chosen such that the two-thirds dealiasing rule \cite{Patterson1971} shall not affect the large scales for any of these adopted grids. With these numerical experiments, it is expected that the sub-Kolmogorov scales should be presented in the $N=128$ case, while certainly not in the $N=64$ case due to the limitation of resolution.

\begin{table*}
\begin{center}
\small
\begin{tabular*}{0.7\textwidth}{@{\extracolsep{\fill}}cccccccc}
\hline
Reso. &$Re_{\lambda}$ &$\nu$ &$\Delta t$ &$\epsilon$ &$\eta$ &$k_{max} \eta$ &$\eta/h_{DNS}$ \\ \hline
$64^3$  &31 &0.04 &0.002 &0.5 &0.106 &2.26 &1.08\\ \hline
$128^3$  &31 &0.04 &0.002 &0.5 &0.106 &4.52 &2.16\\ \hline
\end{tabular*}
\normalsize
\caption{The key numerical parameters for the sub-Kolmogorov recovery test.}\label{tab3}
\end{center}
\end{table*}

To test the sub-Kolmogorov-scale recovery using DA, a new simulation is run using a $N=128$ grid, with a randomly generated initial condition which is `erroneous' with respect to the reference field. Two reference data sources for the DA process are adopted, namely the $N=128$ and the $N=64$ solutions. The former case corresponds to the usually adopted DA procedure, i.e. the large-scale data come from the reference field generated using the same grid resolution, which is known to be viable as long as the large-scale data are sufficient ($k_a>k_c$). In the second case, however, the large-scale reference data are generated using a coarser grid ($N=64$). How such a DA using a coarse-grid solution would behave constitutes the central question of the current section. Both TCDA and TSDA are tested. 

\begin{figure}\centering
\includegraphics[width=.55\textwidth]{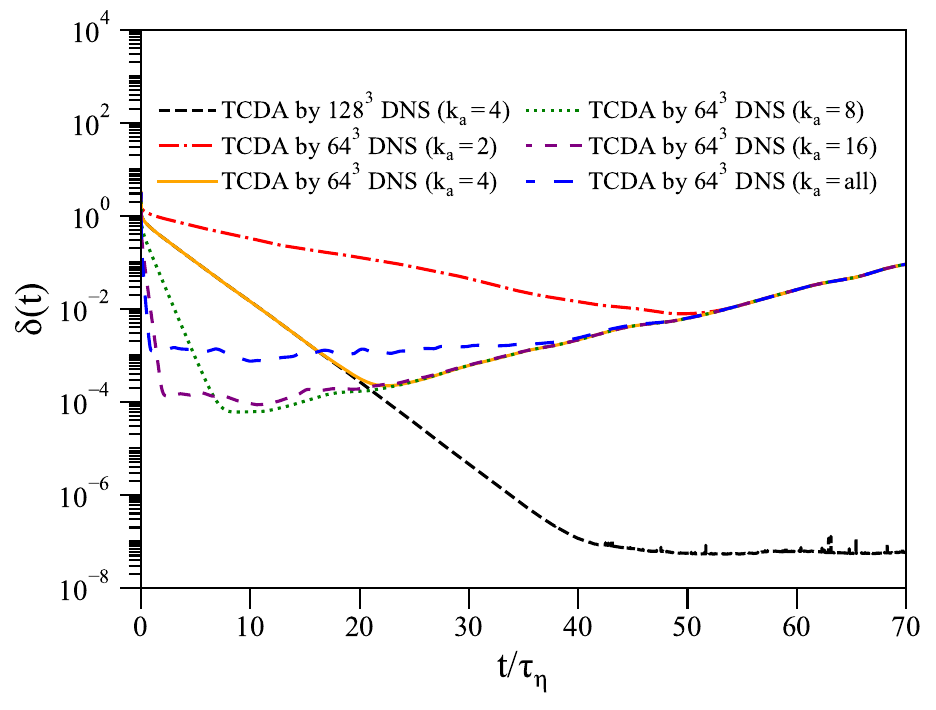}\hspace{-0.10in}
 \caption{The temporal evolution of the error magnitude of the TCDA-based sub-Kolmogorov recovery test.}\label{fig17}
\end{figure}

The evolution of errors using TCDA is shown in Fig.~\ref{fig17}. As expected, the TCDA-based simulation using the $N=128$ DNS easily converges to the reference field at $k_a=4$. Indeed, $k_a \eta > 0.2$ requires only $k_a>2$. In contrast, the cases using the $N=64$ DNS as reference data turn out unsuccessful regardless of the amount of large scales being used. In all the cases using the $N=64$ as reference data, the errors invariably drop initially but increases with time after reaching a minimum. This is somehow expected since the large-scale information obtained from the coarsened $N=64$ grid is not the `correct' large-scale information which in rigorous sense can only be generated using the same resolution at $N=128$.

As observed in Fig.~\ref{fig17}, recovering the sub-Kolmogorov scales using Kolmogorov-scale DNS is not practical at least for the current isotropic turbulence. Nevertheless, the errors still decrease for some time initially since large scales contain less error during the initial stage. Thus, TCDA would still function for some time even in the presence of some propagated errors from the small scales as discussed in Section III. This is also reflected in Fig.~\ref{fig17} that the two $k_a=4$ curves for TCDA using the $N=128$ and the $N=64$ DNS initially overlap. However, as the large-scale errors gradually grow in the $N=64$ case, the corresponding data assimilation fails with two curves finally going separated paths. Meanwhile, all the curves of errors in the failed cases eventually merge into one single curve since the reference data comes from the same resolution. As such, it is not surprising that they all reach the same attractor eventually.

\begin{figure}\centering
\includegraphics[width=.55\textwidth]{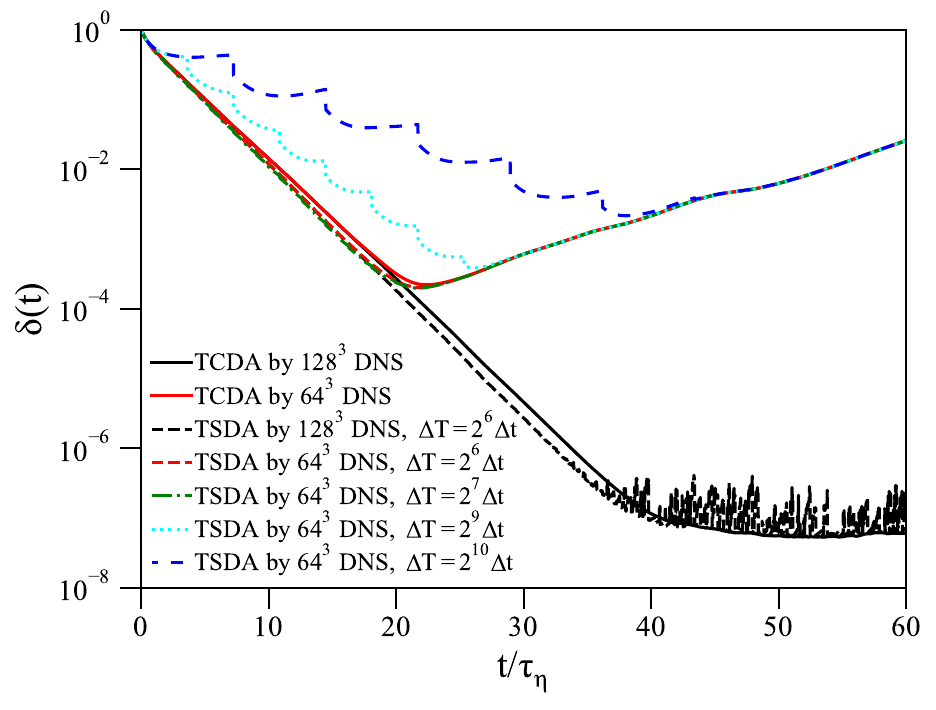}\hspace{-0.10in}
 \caption{The temporal evolution of the error magnitude of the TSDA-based sub-Kolmogorov recovery test for $k_a=4$.}\label{fig18}
\end{figure}

\begin{figure}\centering
\includegraphics[width=.49\textwidth]{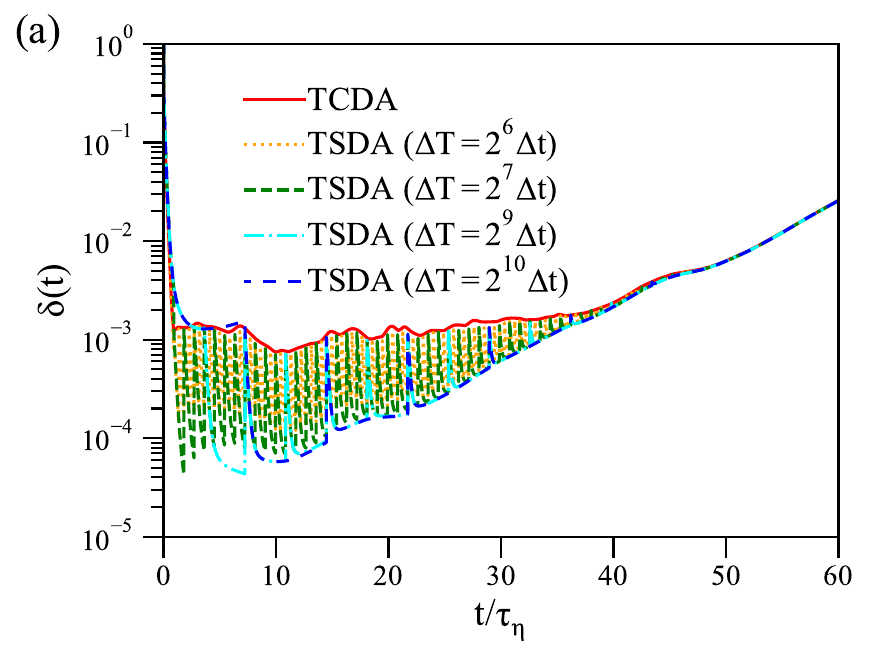}\vspace{-0.10in}
\includegraphics[width=.49\textwidth]{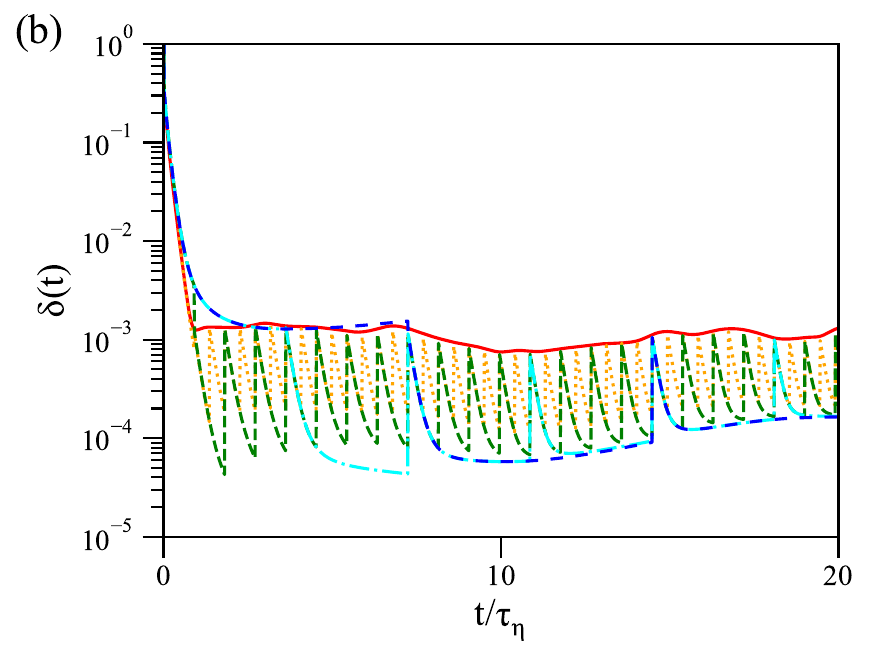}\vspace{-0.10in}
 \caption{The temporal evolution of the error magnitude of the TSDA-based sub-Kolmogorov recovery test for $k_a=all$: (a) $0<t<60 \tau_{\eta}$; (b) $0<t<20 \tau_{\eta}$.}\label{fig19}
\end{figure}

The error evolution in sub-Kolmogorov recovery using TSDA is displayed in Fig.~\ref{fig18} for $k_a=4$, and in Fig.~\ref{fig19} for $k_a=all$ (i.e. all the wavenumbers are used). As can be seen in Fig.~\ref{fig18}, when the reference data source comes from the $N=128$ DNS, TSDA (with $\Delta T=2^6 \Delta t$) also successfully reduces the errors to the machine level in the sub-Kolmogorov recovery. The error decaying rate is even slightly larger than the corresponding TCDA. However, when the reference data source comes from the $N=64$ DNS, TSDA also cannot reconstruct the sub-Kolmogorov structures due to the corruption of the large scales with time. Also, more errors are introduced for $k_a=all$ compared to $k_a=4$ when the reference data is taken from the $N=64$ DNS. Nevertheless, in Fig.~\ref{fig19}a and its zoom-in view Fig.~\ref{fig19}b, it is interesting to observe that TSDA is able to keep the errors below the TCDA level by absorbing data less frequently when $k_a=all$, even though it eventually merges with TCDA due to the growth of errors in the reference field. Indeed, if the reference field becomes completely different from the `true' field with time, the data assimilation would eventually fail regardless of whether TCDA or TSDA is adopted. 

The forgoing results in this section seems to suggest that DA is not a viable method for recovering the sub-Kolmogorov scales using the Kolmogorov-scale solutions. However, one should also be prudent before questioning the validity of the widely adopted DNS resolution due to the limitations in the isotropic turbulence. We recall that for isotropic turbulence, the commonly adopted periodical boundary condition in effect has very little restrictions on the large scales except for the spatial periodicity. Meanwhile, the forcings on the two largest wavenumbers control only the large-scale energy instead of fluid velocity. In this case, the large scales can freely evolve and have more vulnerability to be affected by small-scale perturbations. In contrast, if a flow is bounded by a solid wall, the flow structures probably do not freely evolve without control (at least in the immediate vicinity of the wall). Such issues shall be pursued in future works.

%%%%%%%%%%%%%%%%%%%%%%%%%%%%%%%%%%%%%%%%%%%%%%%%%%%%%%%%%%%%%%%%%%%%%%%%%%%%%%%%%%%%%%%%%%%%%%%%%%%%%%%%%%%%%%%%%%%%%%%%%%%%%%%%%%%%%%%%%%%%%%%%%%%%%
\section{Conclusions}

In the present study, a temporally sparse data assimilation (TSDA) strategy for the reconstruction of small-scale structures of incompressible homogeneous isotropic turbulence (HIT) is proposed. Compared to the previously proposed temporally continuous data assimilation (TCDA), TSDA has significantly reduced the amount of required data while the accuracy is still maintained, or even slightly better in the presence of non-negligible large-scale errors.

Assimilation time steps $\Delta T$ ranging from $2^1$ to $2^{10} \Delta t$ are examined. It is shown that the assimilation time step for TSDA can be at least one order larger than that for TCDA while the performance of TSDA is at least the same as TCDA. Consequently, the amount of required data for TSDA can be significantly reduced compared to TCDA.

To explore the mechanism of TSDA, the one-step data assimilation (ODA) is analyzed through both the evolution of total error magnitude and the error spectrum. For ODA, the error is observed to decrease initially, followed by a exponential growth after hitting a minimum. The initial error decaying rate of ODA is very close to that of TCDA, but the decaying of error gradually slows down. Meanwhile, the behavior of the exponential growth coefficient for the error after it reaches the minimum is found to be consistent with the behavior of the largest Lyapunov exponent.

A detailed comparison is carried out between ODA and TCDA in terms of their error spectra. It is shown that the relaxation effect for the errors above the assimilation wavenumber $k_a$ is responsible for the error decay in ODA. Meanwhile, the errors contained in the large scales can propagate into small scales and make the high-wavenumber ($k>k_a$) error noise decay slower with TCDA than TSDA. This is further confirmed by artificially incorporating different levels of errors in the reference velocity field. The advantage of TSDA is found to grow with the increase of the incorporated errors. Hence, it is potentially more beneficial to adopt TSDA if the reference data contains non-negligible errors.

For a higher Reynolds number case, it is found that, some of the successful assimilation cases experience a transition period before the exponential decay of errors. This can be caused by the fact that, at higher Reynolds number, the energy containing range is more separated from the dissipation range. As a result, for a $k_a$ that is sufficient for the small-scale reconstruction but relatively small, the errors in the far dissipation range only start to drop exponentially after the errors in the intermediate range have sufficiently decreased.

Finally, the possibility of recovering sub-Kolmogorov-scale structure using large-scale information from solutions of Kolmogorov-scale grid is investigated using both TCDA and TSDA. The results show that it is hard to recover the sub-Kolmogorov scales due to the freedom of large scales of HIT. Further investigations on the sub-Kolmogorov influence with more constrained boundary conditions shall be pursued in future works.

\begin{acknowledgments}
This work was supported by the National Natural Science Foundation of China (NSFC Grants No. 91952104, No. 92052301, No. 12172161, and No. 91752201), by the National Numerical Windtunnel Project (No. NNW2019ZT1-A04), by the Shenzhen Science and Technology Program (Grants No. KQTD20180411143441009), by Key Special Project for Introduced Talents Team of Southern Marine Science and Engineering Guangdong Laboratory (Guangzhou) (Grant No. GML2019ZD0103), and by Department of Science and Technology of Guangdong Province (No. 2020B1212030001). This work was also supported by Center for Computational Science and Engineering of Southern University of Science and Technology, and by National Center for Applied Mathematics Shenzhen (NCAMS).
\end{acknowledgments}

\section*{DATA AVAILABILITY}

The data that support the findings of this study are available from the corresponding author upon reasonable request.
%\nocite{*}

\appendix

\section{The influence of grid resolution}

\begin{figure}[H]\centering
\includegraphics[width=.49\textwidth]{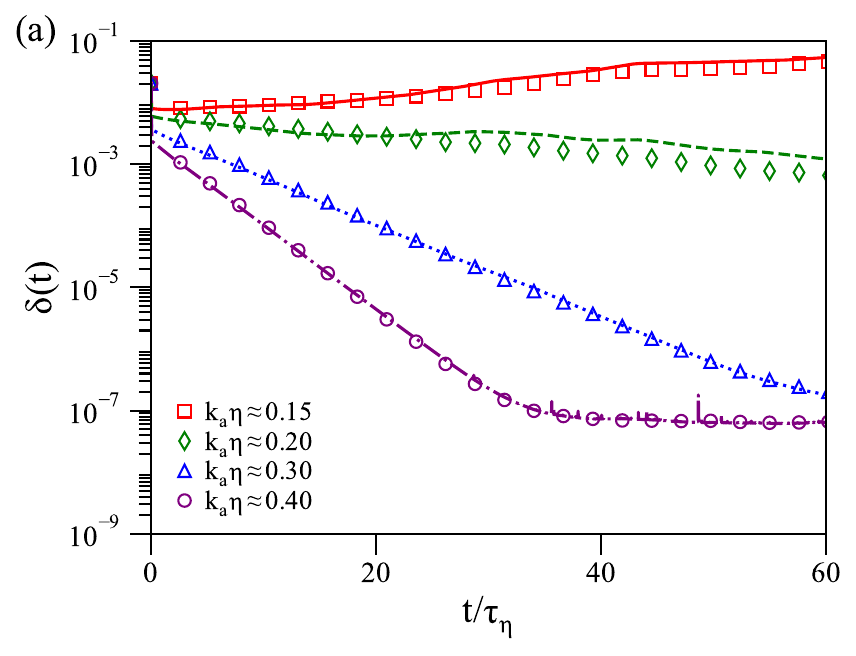}\vspace{-0.10in}
\includegraphics[width=.49\textwidth]{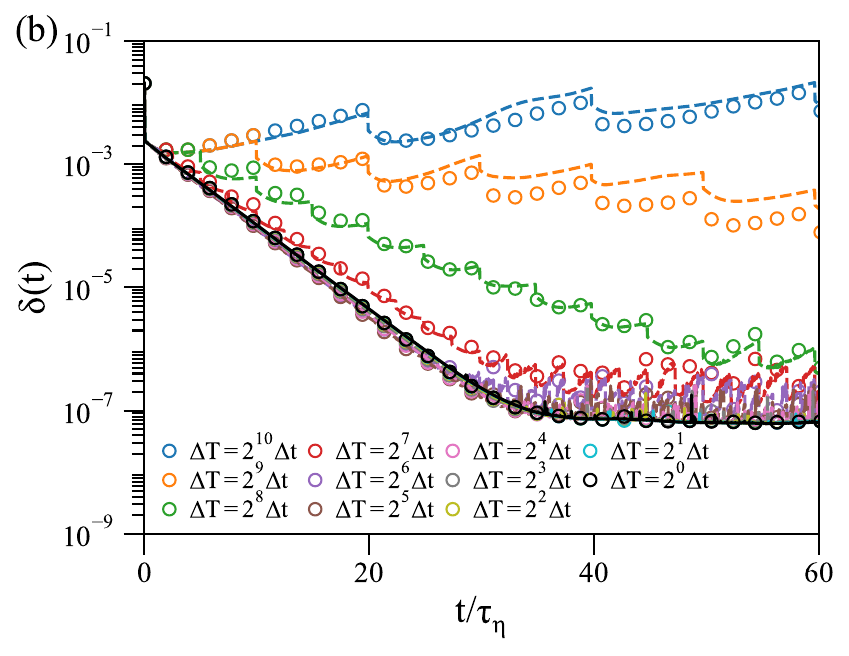}\vspace{-0.10in}
 \caption{The error magnitude evolutions for the TCDA and TSDA of the DNS at $Re_{\lambda}=60$ for different grid resolutions ($k_{max} \eta$). Here the markers represent the results for $k_{max} \eta \approx 2$, and the lines with the same colors as the markers represent the corresponding results for $k_{max} \eta \approx 1$: (a) TCDA; (b) TSDA with $k_a \eta = 0.4$.}\label{fig20}
\end{figure}

In this appendix, we compare the TCDA and TSDA results obtained using $k_{max} \eta \approx 2$ ($N=128$) and $k_{max} \eta \approx 1$ ($N=64$) for the data assimilation of the DNS at $Re_{\lambda}=60$. As can be seen in Fig.~\ref{fig20}a, the difference between the TCDA results using the two grid resolutions are quite small. Therefore, $k_{max} \eta \approx 1$ is sufficient for TCDA. This has been also confirmed by Yoshida \emph{et al.}\cite{Yoshida2005} The results for TSDA (Fig.~\ref{fig20}b) using the two grid resolutions are also very close, except for the cases in which the assimilation step is too large (e.g. $\Delta T=10^9 \Delta t$ and $10^{10} \Delta t$). Here, the value for $\Delta t$ is consistent with that used by the $N=64$ DNS given in Table~\ref{tab1}. In conclusion, the performances of TSDA (as well as that of TCDA) are not much affected by the extra degrees of freedom induced by the finer grid resolution. 

\section{The case of initial error induced by a small phase shift}
\begin{figure}\centering
\includegraphics[width=.49\textwidth]{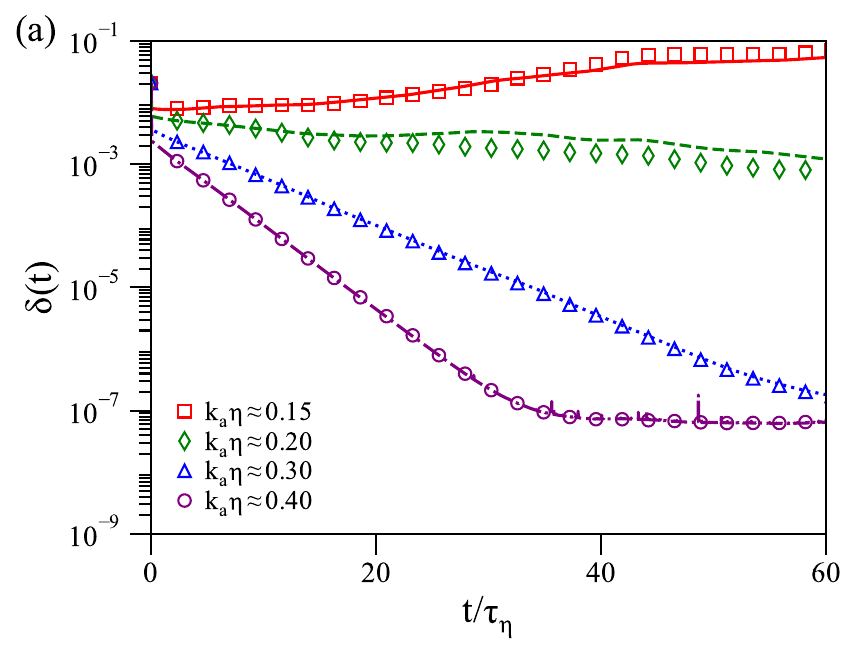}\vspace{-0.10in}
\includegraphics[width=.49\textwidth]{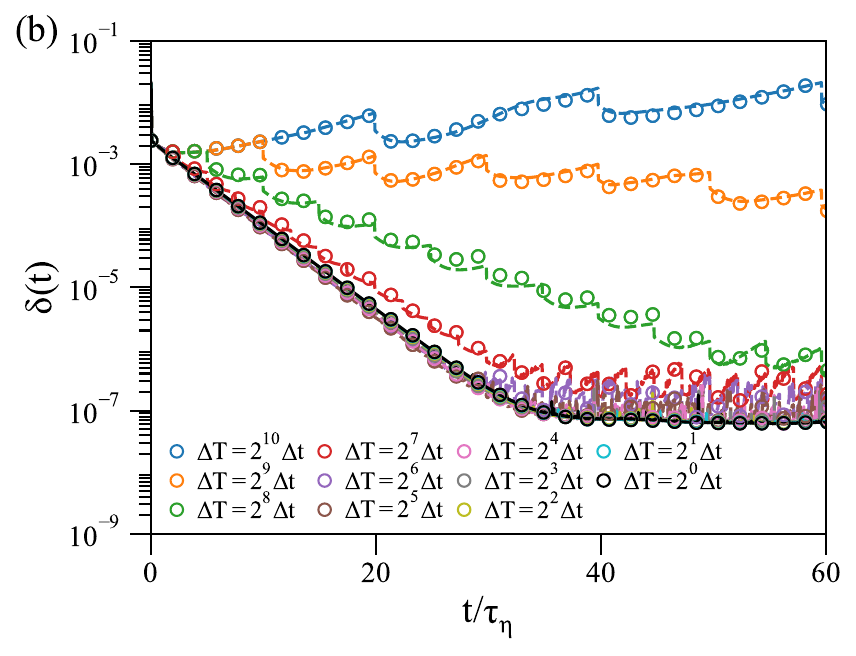}\vspace{-0.10in}
 \caption{The error magnitude evolutions for the TCDA and TSDA of the DNS at $Re_{\lambda}=60$ in the case of a small phase-shift perturbation in the initial Fourier modes. Here the lines with the same colors as the markers represent the corresponding results for the case of small magnitude perturbation in the initial Fourier modes: (a) TCDA; (b) TSDA with $k_a \eta = 0.4$.}\label{fig21}
\end{figure}

In this appendix, we test the data assimilation for the DNS at $Re_{\lambda}=60$ in which the initial perturbation is induced by a small phase shift in the Fourier modes, namely
\begin{equation}
  \mathbf{\hat u}(\mathbf{k},t_0)=(1+\varepsilon ~i)\mathbf{\hat u}^{ref}(\mathbf{k},t_0),
  \label{perturbi}
\end{equation}
where $i=\sqrt{-1}$, and $\varepsilon=10^{-2}$ in consistency with the magnitude perturbation in Section III. The evolutions of the error magnitude for TCDA and TSDA are shown in Figs.~\ref{fig21}a and~\ref{fig21}b, respectively. In Fig.~\ref{fig21}b, the TSDA results are for the $k_a \eta = 0.4$ case. Also shown in the figure are the corresponding results for the cases in which the initial Fourier modes are perturbed in the magnitude as given by Eq.~(\ref{perturb}). As can be seen, the differences between the results generated using the two types of perturbations are quite small for both TCDA and TSDA. Meanwhile, as shown in Fig.~\ref{fig21}b, TSDA can also achieve a similar performance as TCDA with $\Delta T$ at least one order larger than $\Delta t$.  

\section{The case of double-precision solver}
\begin{figure}\centering
\includegraphics[width=.49\textwidth]{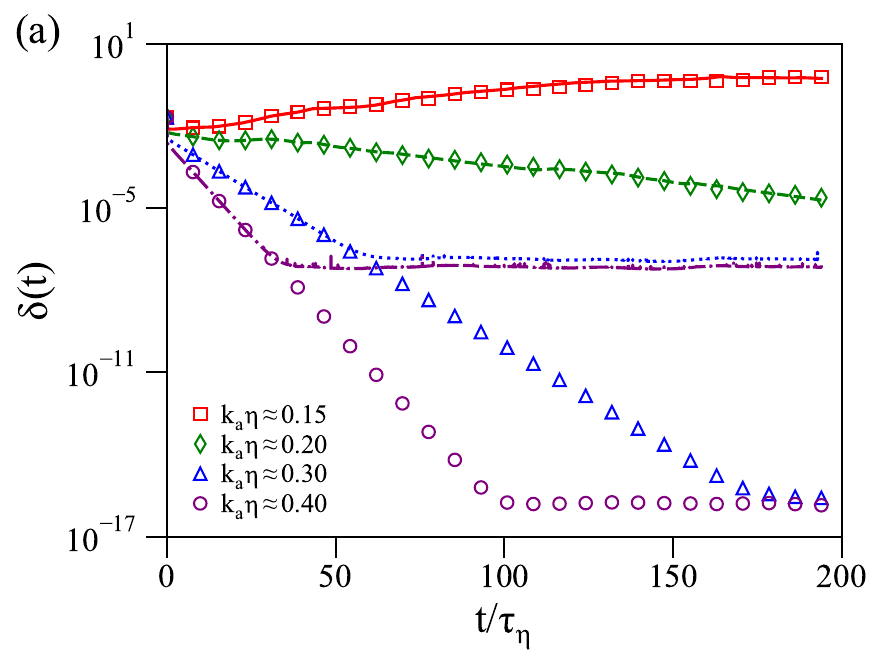}\vspace{-0.10in}
\includegraphics[width=.49\textwidth]{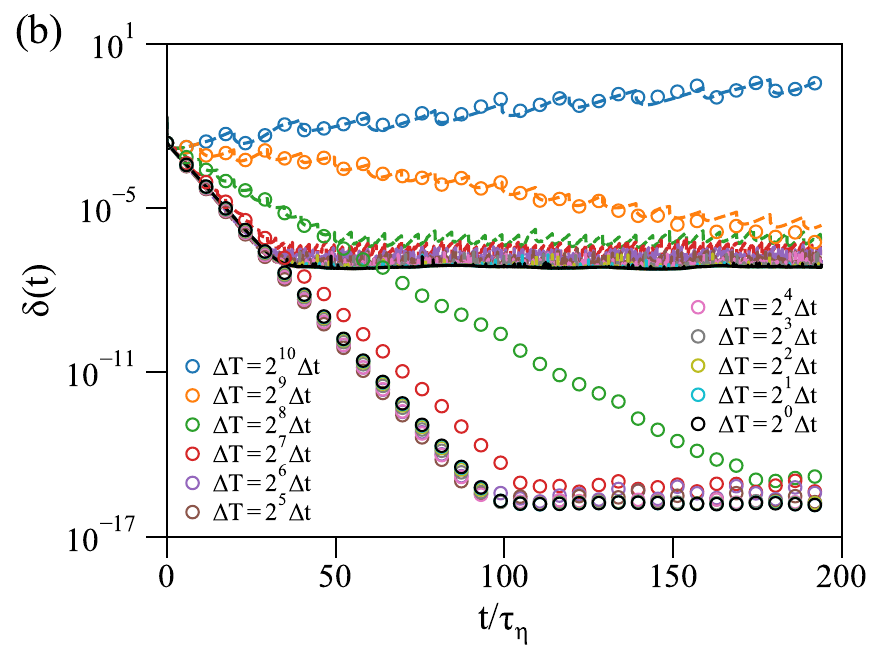}\vspace{-0.10in}
 \caption{The error magnitude evolutions for the TCDA and TSDA of the DNS at $Re_{\lambda}=60$ in the case of double-precision solver. Here the lines with the same colors as the markers represent the corresponding results for the case of single-precision solver: (a) TCDA; (b) TSDA with $k_a \eta = 0.4$.}\label{fig22}
\end{figure}

In this appendix, we compare the results obtained using the single- and double-precision solvers for the data assimilation of the DNS at $Re_{\lambda}=60$. In general, the error magnitude can be reduced to the order of $10^{-6}$ with a successful data assimilation using a single-precision solver, which is confirmed in the current work and also by Lalescu \emph{et al.}\cite{Lalescu2013} On the other hand, with a double-precision solver, the corresponding error magnitude should be reduced to $10^{-15}$. In Figs.~\ref{fig22}a and \ref{fig22}b, the results for error evolution generated using the double-precision solver are shown for TCDA and TSDA, respectively, along with the corresponding results generated using the single-precision solver. As expected, the double-precision solver reduces the error to a level below $10^{-15}$. Except for this, the results generated using the two solvers agree very well for both TCDA and TSDA, and the round-off error shall not influence much the presented results as well as the advantage of TSDA. In fact, as shown in Section III, the advantage of TSDA increases with the errors contained in the reference data.

\end{document}